\documentclass[10pt]{article}

\usepackage{geometry}  
\usepackage[english]{babel}
\usepackage[utf8x]{inputenc}
\usepackage[T1]{fontenc}

\usepackage[dvipsnames]{xcolor}
\usepackage{paralist}
\usepackage{graphicx}
\usepackage{subcaption}
\usepackage{longtable} 
\usepackage{multirow}
\usepackage{listings}
\usepackage{makecell}
\usepackage{array}
\usepackage{float}
\usepackage{dsfont}
\usepackage{rotating}
\usepackage{booktabs}
\usepackage{enumerate}
\usepackage{tikz}
\usepackage{pgf}
\usepackage{xcolor}
\usepackage[title]{appendix}

\usepackage{amsmath}
\usepackage{amssymb}
\usepackage{amsthm}
\usepackage{bm}
\usepackage{bbm}
\usepackage{mathtools}
\usepackage{mathrsfs}
\usepackage{algorithm}
\usepackage{algorithmic}

\usepackage{natbib}
\usepackage[
  colorlinks,
  citecolor=blue,
  linkcolor=red,
  anchorcolor=red,
  urlcolor=blue
]{hyperref}
\mathtoolsset{showonlyrefs}
\usepackage{authblk}
\usepackage{todonotes}
\usepackage{hyperref}
\theoremstyle{plain}

\newtheorem{definition}{Definition}
\newtheorem{theorem}{Theorem}

\newtheorem{remark}{Remark}
\newtheorem{assumption}{Assumption}

\newcommand{\ind}{\mathbbm{1}}

\newcommand{\indep}{\rotatebox[origin=c]{90}{$\models$}}

\usepackage{xspace}

\newcommand{\given}{{\,|\,}}
\newcommand{\biggiven}{\,\big{|}\,}
\newcommand{\Biggiven}{\,\Big{|}\,}
\newcommand{\bigggiven}{\,\bigg{|}\,}
\newcommand{\Bigggiven}{\,\Bigg{|}\,}
\def\##1\#{\begin{align}#1\end{align}}
\def\$#1\${\begin{align*}#1\end{align*}}

\definecolor{myblue}{rgb}{.8, .8, 1}
\definecolor{mathblue}{rgb}{0.2472, 0.24, 0.6} 
\definecolor{mathred}{rgb}{0.6, 0.24, 0.442893}
\definecolor{mathyellow}{rgb}{0.6, 0.547014, 0.24}

\newcommand{\calA}{{\mathcal{A}}}

\newcommand{\calD}{{\mathcal{D}}}
\newcommand{\calE}{{\mathcal{E}}}

\newcommand{\calI}{{\mathcal{I}}}

\newcommand{\calN}{{\mathcal{N}}}

\newcommand{\calX}{{\mathcal{X}}}

\newcommand{\EE}{\mathbb{E}}
\newcommand{\PP}{\mathbb{P}}
\newcommand{\RR}{\mathbb{R}}

\long\def\comment#1{}

\let\hat\widehat
\let\tilde\widetilde

\def \defn {\,:=\,}
\def \eps {\varepsilon}

\usepackage{hyperref}
\makeatletter
\newcommand{\printfnsymbol}[1]{%
  \textsuperscript{\@fnsymbol{#1}}%
}
\makeatother

\usepackage{xcolor}

\title{Conformalized survival analysis with adaptive cutoffs}
\author[1]{Yu Gui\thanks{Authors listed alphabetically.}}
\author[1]{Rohan Hore\printfnsymbol{1}}
\author[2]{Zhimei Ren\printfnsymbol{1}}
\author[1]{Rina Foygel Barber}
\affil[1]{Department of Statistics, University of Chicago}
\affil[2]{Department of Statistics and Data Science, The Wharton School, University of Pennsylvania}
\date{\today}

\begin{document}

\maketitle

\begin{abstract} 
This paper introduces an assumption-lean method that constructs valid and 
efficient lower predictive bounds (LPBs) for survival times
with censored data. We build on recent work by~\citet{candes2021conformalized}, 
whose approach first subsets the
data to discard any data points with early censoring times, and then uses
a reweighting technique (namely, weighted conformal inference~\citep{tibshirani2019conformal})
to correct for the distribution shift introduced by this subsetting
procedure. 

For our new method, instead
of constraining to a fixed threshold for the censoring time when subsetting the data,
we allow for a covariate-dependent and data-adaptive
subsetting step, which is better able to capture the heterogeneity of the 
censoring mechanism. As a result, our method can lead to LPBs that are 
less conservative and give more accurate information.
We show that in the Type I right-censoring setting, if 
either of the censoring mechanism or the conditional 
quantile of survival time is well estimated, our 
proposed procedure achieves nearly exact
marginal coverage, where in the latter case we 
additionally have approximate conditional coverage.
We evaluate the validity and efficiency of our proposed algorithm 
in numerical experiments, illustrating its 
advantage when compared with other competing methods. 
Finally, our method is applied to a real dataset to generate LPBs
for users' active times on a mobile app.
\end{abstract}

\section{Introduction}
Survival analysis lies at the core of many important questions in clinical trials~\citep{fleming2000survival,singh2011survival}, ecology~\citep{muenchow1986ecological}, and other applied fields.
In particular, one important problem is that of studying the behavior of  survival time $T$, and how it relates to other features of the data, which we denote by a potentially high-dimensional feature vector $X$. Modeling the association between $X$ and $T$
can in turn play a crucial role in enabling more useful and reliable policy making. 
The major challenge is that these survival times are only partially observed due to censoring \citep{leung1997censoring}, which makes the statistical analysis quite non-routine---we are
only able to observe the survival time $T$
if it occurs no later than some censoring time $C$. For example, $T$ may be the survival time of a patient 
(measured as time since diagnosis), which may be censored at a time $C$ that denotes the endpoint of the study that follows the patient.

One of many goals of survival analysis is to infer 
the survival function---the probability of survival
beyond a given time---given the censored data. The
Kaplan-Meier curve \citep{kaplan1958nonparametric} 
can produce such inferences for sub-population with 
a particular covariate structure while making no 
assumption on the distribution of survival times,
but it requires sufficiently many events in each subgroup \citep{kalbfleisch2011statistical}. 
This assumption is 
no longer realistic in the modern era of big data, where
with the ever-increasing ability to collect and store data, 
we can have access to a large number of (potentially continuous)
covariates.

Over the years,  
many tools have been developed to cope with
such high dimensionality, offering
estimation of  the conditional survival function.
One popular example in that line is the Cox model 
which posits a proportional hazard model:
an unspecified non-parametric baseline is modified
via a parametric model describing how the hazard
varies in response to explanatory 
covariates~\citep{cox1972regression,breslow1975analysis}.  
Other popular parametric approaches include the accelerated
failure time (AFT) model~\citep{cox1972regression,wei1992accelerated} 
and the proportional odds model~\citep{murphy1997maximum,harrell2015regression}.
More recently, we have witnessed more complex survival
analysis methods that are based on machine learning/deep 
learning~\citep{faraggi1995neural,tibshirani1997lasso,
gui2005penalized,katzman2016deep,lao2017deep,wang2019machine,li2020censored}. 
Despite the success of these methods in many areas,
it remains largely unclear how to provide reliable
uncertainty quantification for these methods. This is
mainly because they posit model assumptions that are 
hard to verify and/or the algorithms themselves are
too complicated
to be analyzed. For these reasons, it is desirable to 
find a more assumption-lean or distribution-free approach
towards reliable inference in survival analysis.

The recent work of \citet{candes2021conformalized}  proposes such an approach (which 
we will describe in detail below). As the target of inference,
they propose computing a $100(1-\alpha)\%$ lower 
prediction bound (LPB) for the survival time of a patient/unit, 
where $\alpha$ is a pre-specified level; it
means that the patient/unit is expected to survive 
beyond this predicted time with at least $100(1-\alpha)\%$
probability.
The LPB is used to provide
a summary of what we can infer about the individual's survival time given available
data, and it is important to note that the LPB can be low when
either the true survival time is low or there is not enough information
for us to get an informative lower bound; in other words, insufficient data 
should not lead to an invalid claim, but instead may lead to a less informative
output.

\subsection{Defining the lower prediction bound (LBP)}
Let $X\in \calX$ denote the 
covariate vector, $T\in \RR_{\ge 0}$ the survival time,
and $C \in \RR_{\ge 0}$ the censoring time. Under censoring,
the survival time $T$ is observed only if it occurs before the censoring time $C$.
In other words, while the features $X$ and the censoring time $C$ are both observed,
the survival time is observed only indirectly, via the
censored survival time as $\tilde{T} = \min(T,C)$ (which may not be equal to $T$). 

We now
give the formal definition
of a marginally calibrated LPB.
Throughout, for a joint distribution $P$ on $(X,C,T)$, we will write $P_X$, $P_{(X,T)}$, $P_{(X,\Tilde{T})}$, etc, to denote
the corresponding marginal distributions, and $P_{C|X}$, $P_{T|X}$, $P_{\Tilde{T}|X}$, etc, to denote the corresponding conditional
distributions.

\begin{definition}[Marginally calibrated LPB]
\label{def:calibrated_LPB}
Let $(X_i,C_i,T_i)\stackrel{\textnormal{iid}}{\sim} P$ for data points $i=1,\dots,n$, and let $\hat{L}$ be a function of the observed
data $\calD=\{(X_i,C_i,\Tilde{T}_i):1\leq i\leq n\}$, where $\Tilde{T}_i=\min(T_i,C_i)$ is the censored
survival time. Then we say that $\hat{L}$ is a marginally calibrated 
LPB at level $1-\alpha$ if it satisfies
\begin{equation}\label{eqn:marginal_LBP_def}\mathbb{P}_{(X,T) \sim P}\big(T\geq \hat{L}(X)\big)\geq 1-\alpha ,\end{equation}
where this probability is taken with respect to both the available data $\calD$
and a new data point $(X,T)\sim P_{(X,T)}$.
\end{definition}

The marginally calibrated LPB provides guarantee in an average
sense---that is, over all the possible draws of the data, the 
coverage of the LPBs is guaranteed. However, in practical settings,
we may be more concerned about the coverage guarantee 
we can obtain given the data at hand. There, the  probably approximately correct 
(PAC)-type LPB  defined below can be more informative~\citep[see also][]{vovk2012conditional,bates2021distribution,angelopoulos2021learn,jin2021sensitivity}. 

\begin{definition}[PAC-type LPB]
\label{def:pac_LPB}
Under the same notation as Definition~\ref{def:calibrated_LPB},
we say that $\hat{L}$ is a
PAC-type LPB at level $\alpha$ with tolerance 
$\delta$ if, with probabilty at least
$1-\delta$ over the draw of $\calD$,
$$\mathbb{P}_{(X,T)\sim P}\big(T\geq \hat{L}(X) \given \calD\big)\geq 1-\alpha $$
where the probability is now taken with respect to a new data point $(X,T)\sim P_{(X,T)}$.
\end{definition}

Throughout, we adopt the {\em conditionally independent censoring} assumption.
\begin{assumption}[Conditionally independent censoring]
\label{assumption:conditionally_independent_censoring}
The joint distribution $P$ of $(X,C,T)$ satisfies $C \,\indep \, T \given X$.
\end{assumption}
\noindent This assumption  is standard
in the survival analysis literature, in order to ensure identifiability of 
the problem~\citep[see e.g.,][]{kalbfleisch2011statistical}.

\subsection{An initial approach: inference on the censored survival time}\label{sec:intro_censored}
As discussed by \citet{candes2021conformalized},
since the censored survival time $\Tilde{T}$ cannot be larger than $T$ by definition,
any valid lower bound on $\Tilde{T}$ is trivially a lower bound on $T$. In other words,
if an estimated lower bound $\hat{L}$ satisfies
$\mathbb{P}_{(X,C,T)\sim P}\big(\Tilde{T}\geq \hat{L}(X)\big)\geq 1-\alpha$,
then trivially Definition~\ref{def:calibrated_LPB} is satisfied and so $\hat{L}$ is a marginally
calibrated LPB.
(Similarly, if $\mathbb{P}\big(\Tilde{T}\geq \hat{L}(X)\given 
\calD\big)\geq 1-\alpha $ with probability at least $1-\delta$,
then by Definition~\ref{def:pac_LPB} $\hat{L}$ is a PAC-type LPB.)
Since the censored survival time $\tilde{T}$ can be observed in the dataset at hand
(and so $\hat{L}$ can be constructed to satisfy this property),
this provides a mechanism for providing a valid LPB. 

However, if the censoring time $C$
is often substantially smaller than $T$, then a valid lower bound on $\Tilde{T}$ may 
be extremely conservative as a lower bound on $T$ itself, thus reducing the utility of the constructed LPB.
This suggests that such an approach may not be optimal for most applications.
On the other hand, \citet{candes2021conformalized} prove that, in the absence of any assumptions
on the distribution $P$ on $(X,C,T)$, it is impossible to improve on this type of approach---specifically,
their result \citep[Theorem 1]{candes2021conformalized} proves that,
for any construction $\hat{L}$ that satisfies Definition~\ref{def:calibrated_LPB} universally over
all distributions $P$, $\hat{L}$ must also satisfy $\mathbb{P}\big(\Tilde{T}\geq \hat{L}(X)\big)\geq 1-\alpha $.
This motivates their introduction of an additional assumption, as we describe next.

\subsection{\citet{candes2021conformalized}'s approach: a cutoff on the censoring time}\label{sec:intro_c0}
As described above, constructing an LPB on the censored survival time $\Tilde{T}$ may be too conservative
in applications where the censoring time $C$ is frequently low, leading to censored times $\Tilde{T}$
that are far smaller than the true target of inference $T$. \citet{candes2021conformalized}'s
approach is to avoid this issue by discarding any training data points where $C$ is very low---specifically,
for a constant cutoff $c_0$, they subset the data $\calD$
to keep only data points $(X_i,C_i,\Tilde{T}_i)$ for which $C_i\geq c_0$. After this filtering
step, any lower bound $\hat{L}$ on
the remaining censored survival time $\Tilde{T}$ is no longer necessarily overly conservative,
since the condition $C\geq c_0$ (with a well-chosen $c_0$)
ensures that $\Tilde{T}$ is less likely to be far smaller than $T$.
Thus, we can proceed by constructing an LBP $\hat{L}$ that is a lower bound on $\Tilde{T}$, in this
new training sample.

Of course, we must
then be careful about biasing the results because of this cutoff. In particular, since the event $C\geq c_0$
may be highly dependent on the covariates $X$, 
the remaining data is drawn from a distribution that is different from the target distribution $P$.
To be more precise, writing $P^{\geq c_0}$ to denote the distribution of a data point $(X,C,T)\sim P$ given the event $C\geq c_0$,
we see that the remaining data consists of samples from $P^{\geq c_0}$ while the 
inference goal is to provide coverage under the original distribution $P$. In other words,
we would like to ensure that the marginal coverage bound~\eqref{eqn:marginal_LBP_def} holds,
but calibrating $\hat{L}(\cdot)$ na\"ively 
on the remaining data would instead
only ensure that
$\PP_{(X,T)\sim P^{\geq c_0}}(T \geq \hat{L}(X))\geq 1-\alpha,$
or equivalently,
$\PP_{(X,T,C)\sim P} ( T\geq \hat{L}(X) \given C\geq c_0) \geq 1-\alpha$.

To account for this shift in the distribution, 
\citet{candes2021conformalized} utilize the method of {\em conformal prediction
under covariate shift} \citep{tibshirani2019conformal}, which builds on the 
well-known conformal prediction framework for distribution-free predictive inference \citep{vovk2005algorithmic}.
To do so, they
 additionally assume that we have exact or approximate
knowledge of the dependence of censoring time $C$ on the covariates $X$---that is,
knowledge of $P_{C|X}$, or more specifically, $\PP(C\geq c_0|X)$. 
With this additional information, we can reweight the remaining data points to correct
for the change in distribution---essentially, similarly to inverse propensity score weighting,
 weights $1/\PP_P(C\geq c_0|X)$ can account for the difference between the target distribution $P$
and its filtered version $P^{\geq c_0}$. (Of course, the best value of $c_0$ will depend on the data
distribution, and in practice can be chosen on a training set.)

\subsection{Our approach: the benefits of a covariate-adaptive cutoff}\label{sec:intro_example}
In the method described above, how should the cutoff $c_0$ be chosen? 
The choice of $c_0$ presents a tradeoff: if $c_0$ is chosen to be too small,
then the inequality $\Tilde{T}\leq T$ might be quite loose, and the constructed LPB $\hat{L}$
might still be very conservative even after filtering the data with the cutoff. On the other
hand, if $c_0$ is chosen to be too large, then $\PP_P(C\geq c_0|X)$ may be quite small (at least, for many values of $X$),
leading to a low effective sample size, 
large weights $1/\PP_P(C\geq c_0|X)$ on these data points, and highly unstable behavior.
In fact, it is not always possible to find a constant $c_0$ that yields good 
LPBs, especially in cases when the censoring time varies substantially with 
respect to the covariates $X$---selecting a large value of $c_0$ could cause 
highly unstable LPBs in areas where censoring times are low, whereas selecting 
a small value of $c_0$ leads to conservative LPBs in areas where censoring times 
are actually high. 
To be more specific, think of a simple example where $X\sim \text{Unif}([0,1])$ and 
$C = a\ind\{X \ge \frac{1}{2}\} + b\ind\{X<\frac{1}{2}\}$ with $a \gg b$;
choosing $c_0$ to be greater than $b$ requires dropping half of the data and 
leads to  increased variability; instead, selecting a $c_0\le b$ yields very conservative LPBs for 
$X\ge \frac{1}{2}$.

From the above discussion, we can see that it may be beneficial to allow $c_0$ to depend on $X$.
That is, if $\PP_P(C\geq c_0|X)$ is extremely small then we may need to instead choose a lower value of $c_0$
to avoid high variance, but if $\PP_P(C\geq c_0|X)$ is close to 1 then we can afford to increase the value
of $c_0$, thus avoiding an overly conservative LPB. To illustrate the benefits of this 
more flexible approach, we show a small simulated example.

\begin{figure}[!h]
    \centering
    \begin{minipage}{0.3\textwidth}
    \includegraphics[width=\textwidth]{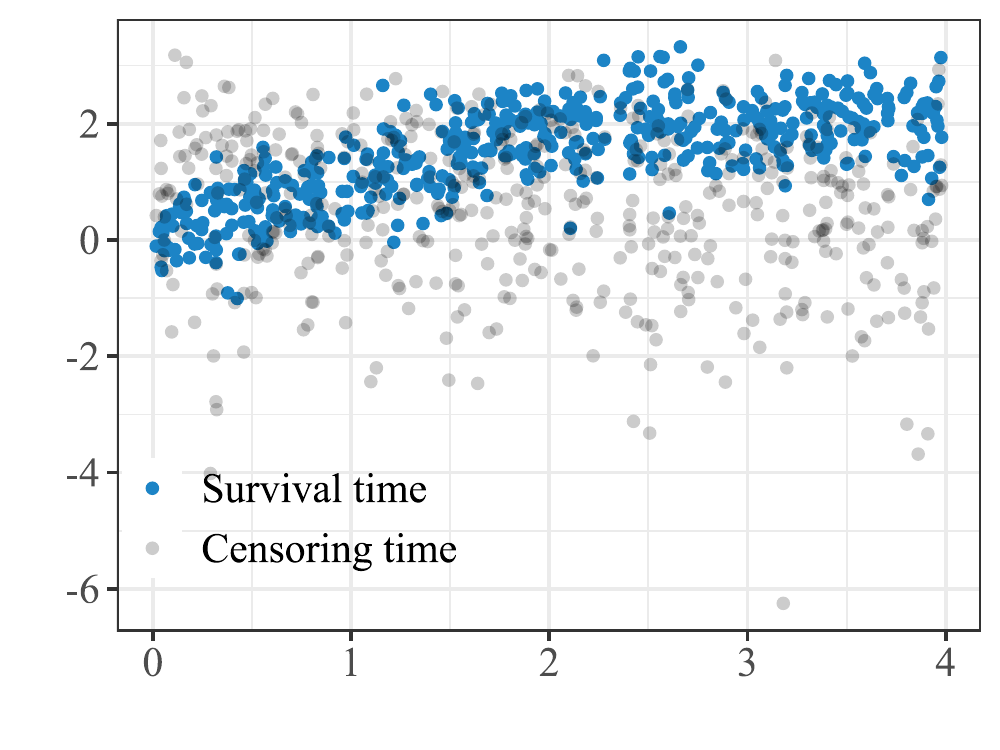}
    \end{minipage}
    \begin{minipage}{0.33\textwidth}
    \includegraphics[width=\textwidth]{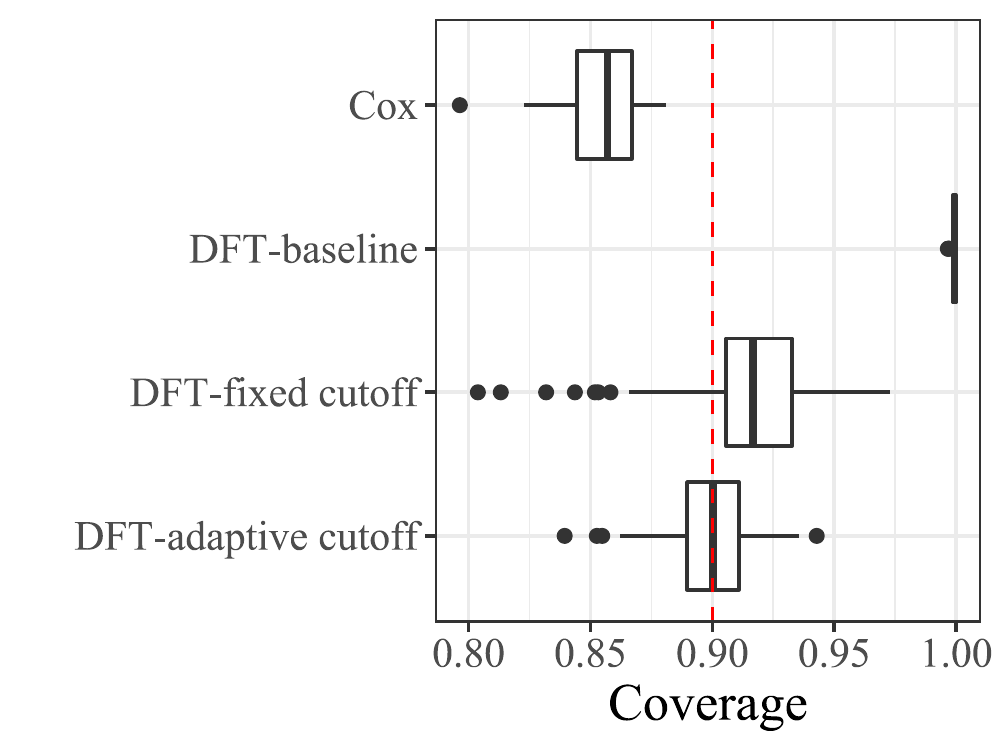}
    \end{minipage}
    \begin{minipage}{0.33\textwidth}
    \includegraphics[width=\textwidth]{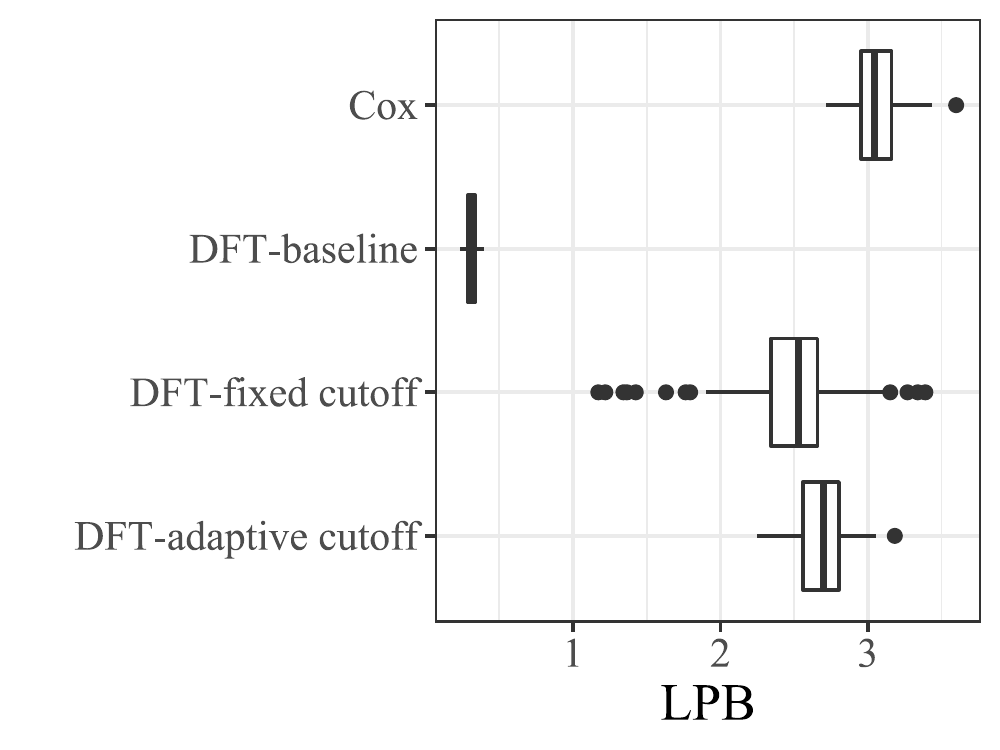}
    \end{minipage}
    \caption{Left: an illustration of the training sample for one trial of the experiment. 
      Middle: boxplot of the coverage rate; the red dashed line corresponds 
      to the target coverage rate $1 - \alpha  = 90\%$. Right: boxplot of the LPBs. The results
      are from $100$ independent trials.}
    \label{fig:covariate_dependence_effect}
\end{figure}

We consider a univariate-covariate case, where $T$ and $C$
depend on $X$ via different models (the details are to be
given in Section~\ref{sec:sims}).
The left panel of Figure~\ref{fig:covariate_dependence_effect}
visualizes (one realization of) the censoring time and survival 
time as functions of the covariate. In this example,  
units with larger values of $X$ tend to have
lower censoring times ($P_{C \given X } = 
\text{Exp}(0.25 + (6+x)/100)$), and thus we should choose a 
lower value of $c_0$ to avoid
high variance (i.e., to avoid overly large weights $1/\PP_P(C\geq c_0|X)$; 
units with smaller values of $X$, on the other hand, 
tend to have larger values of $C$ and so we can afford to increase
the value of $c_0$, leading to a less conservative LPB.

From this model, $n = 2,\!000$ independent samples are generated. 
We compare the baseline method introduced in Section~\ref{sec:intro_censored} (referred to as
DFT-baseline, where DFT is short for ``distribution-free (LPB) for $T$''), \citet{candes2021conformalized}'s fixed cutoff method (referred to as DFT-fixed cutoff),
our new adaptive cutoff method (referred to as DFT-adaptive cutoff and to be defined shortly), and the Cox parametric
model. 
The generated LPBs are then evaluated with an independent dataset of 
$5,\!000$ test samples, and we display the coverage rate and the resulting LPB
in the middle and right panels of Figure~\ref{fig:covariate_dependence_effect}, respectively,
with results gathered from $100$ independent trials.
The parametric method fails to cover
the true survival time with desired probability;
the baseline method and, to a lesser extent, the fixed cutoff method are conservative in this setting,
returning a low (i.e., less informative) LPB. 
On the other hand, our adaptive cutoff method is able to avoid  under-
or over-coverage;
it achieves essentially the target coverage rate and returns a
higher (i.e., more precise) LPB.


\section{Background}\label{sec:background}

\subsection{Covering the censored survival time via CQR}
As described in Section~\ref{sec:intro_censored}, it is possible to provide an LPB on $T$ 
with no further assumptions by simply finding a lower bound on the censored survival time $\Tilde{T}\leq T$.
To do so, one approach is to use the Conformalized Quantile Regression (CQR) framework of \cite{romano2019conformalized}.
To begin, we first partition the available $n$ data points into two data sets, a training set $\mathcal{I}_1$ and a calibration set $\mathcal{I}_2$---for instance,
into two sets of size $n/2$. Using the training set, we fit a quantile regression:
$x \mapsto \widehat{q}_\alpha(x)$,
which estimates the conditional $\alpha$-quantile of $T$ given $X$. This may be done using
an arbitrary algorithm, for instance, linear regression or random forests. If this quantile regression were fitted accurately,
then we could simply use $\widehat{q}_\alpha(X)$ as a LPB for $T$---if indeed this is the $\alpha$-quantile of $T\given X$, 
then $T\geq \widehat{q}_\alpha(X)$ holds with probability $1-\alpha$, as desired. However,
due to potential issues of overfitting, model misspecification, etc, we cannot rely on this being the case, and so
the calibration set is then used to correct for any errors in the initial model fitting stage. For each $i\in \mathcal{I}_2$ (the calibration
points), define a score $V_i = \widehat{q}_\alpha(X_i) - \Tilde{T}_i$,
and then define the LPB as
\[\widehat{L}_{\mathrm{baseline}}(X) = \widehat{q}_\alpha(X) - Q_{1-\alpha}\Big(\sum_{i\in\mathcal{I}_2} \frac{1}{1+|\mathcal{I}_2|}\cdot\delta_{V_i} +\frac{1}{1+|\mathcal{I}_2|}\cdot\delta_{+\infty}\Big),\]
where
$Q_{1-\alpha}(\cdot)$ denotes the $(1-\alpha)$-quantile of a distribution and where $\delta_v$ is the point mass at $v$.
The intuition here is that the $Q_{1-\alpha}(...)$ term adds a correction to the original fitted model to ensure
that $\widehat{L}_{\mathrm{baseline}}(X)$ has the right coverage level on the calibration set drawn i.i.d.~from $P$, and will thus have the right coverage
level on a future draw $(X,T)$ from $P_{(X,T)}$ as well.

Note that the resulting value $\widehat{L}_{\mathrm{baseline}}(X)$ may be higher (less conservative) or lower (more conservative) than the initial
fitted model $\widehat{q}_\alpha(X)$, depending on whether the original fitted model $\widehat{q}_\alpha$ is over- or under-covering
on the calibration set. In practice, 
it is likely that we will have undercoverage of the original fitted model 
(see, e.g., the simulation results in~\citet{romano2019conformalized,lei2020conformal,candes2021conformalized} and Figure~\ref{fig:illustration}),
leading to a quantile $Q_{1-\alpha}(...)$ that is positive, and a LPB $\widehat{L}(X)$ that is lower (more conservative) than the original fitted model.

The following result proves that this is a marginally calibrated LPB:
\begin{theorem}[{Adapted from Theorem 1 of~\citet{romano2019conformalized}}]
Suppose $(X_i,C_i,T_i)\stackrel{\textnormal{iid}}{\sim} P$. Then $\widehat{L}_{\mathrm{baseline}}(X)$ is a marginally calibrated LPB at level $1-\alpha$, and moreover, satisfies
$$\mathbb{P}_{(X,T,C)\sim P}\big(\Tilde{T}\geq \widehat{L}_{\mathrm{baseline}}(X)\big)\geq 1-\alpha.$$
\end{theorem}
\noindent Since $\Tilde{T}$ may be often much smaller than $T$ if the censoring is severe, this
result indicates that such an LPB may be quite conservative as a lower bound for $T$. This conservativeness is however 
inescapable without further assumptions---\citet[Theorem 1]{candes2021conformalized} establish
that, under mild conditions, for any marginally calibrated LPB $\widehat{L}$ for the (uncensored) survival time $T$ which is 
valid universally over all distributions $P$ on the data, $\widehat{L}$ must also be an LPB for $\Tilde{T}$
whenever $P_{(C,T)}$ is either discrete or continuous. 

\subsection{Using fixed threshold $c_0$}
Next we give details for \citet{candes2021conformalized}'s proposed method, which uses a fixed threshold
$c_0$ to avoid an overly conservative LPB. As mentioned above, their work shows that, without further assumptions,
it is not possible to improve on the LPB for $\Tilde{T}$; therefore, they make the additional assumption
that the conditional distribution $P_{C|X}$ is known (or is estimated accurately).

As for CQR, their method begins by partitioning the data into a training set $\mathcal{I}_1$ and a calibration
set $\mathcal{I}_2$, and uses the training set to  fit a quantile regression,\footnote{While their proposed method
is defined via a more general construction, here we focus on a single version that is most relevant for comparison
to our own methods.}
$x \mapsto \widehat{q}_\alpha(x)$,
for the conditional $\alpha$-quantile of $T$ given $X$. The cutoff $c_0$ for the censoring time
may also be chosen as a function of the training data. Furthermore, define
$\widehat{w}(x)$ to be an estimate of $1/\PP(C\geq c_0|X=x)$ (or, approximately proportional
to this quantity), also fitted on the training data.

Next, on the calibration set,
we use $c_0$ to filter the data and define $\mathcal{I}_2 ' = \{i\in\mathcal{I}_2 : C_i \geq c_0\}$.
For all these remaining calibration points, note that $\Tilde{T}_i\wedge c_0 = T_i\wedge c_0$ (that is,
$T_i\wedge c_0$ is observed).
We then calculate scores $V_i = \widehat{q}_\alpha(X_i) -T_i\wedge c_0$
for all $i\in\mathcal{I}'_2$, and return the LPB
\[\widehat{L}_{\mathrm{fixed-cutoff}}(X) = \widehat{q}_\alpha(X) - Q_{1-\alpha}\Big(\frac{\sum_{i\in\mathcal{I}'_2} \widehat{w}(X_i) \cdot\delta_{V_i} + \widehat{w}(X)\cdot\delta_{+\infty}}{\sum_{i\in\mathcal{I}'_2} \widehat{w}(X_i) + \widehat{w}(X)}\Big).\]
The intuition here is that the calibration set $\calI_2'$ 
consists of data points drawn from the shifted distribution $P^{\geq c_0}$,
and the likelihood 
ratio between the target distribution $\PP$ and this distribution $\PP^{\ge c_0}$
is $\PP(C\ge c_0)/\PP(C\ge c_0 \given X)$; since $\hat{w}(X)$ 
is an estimate of the likelihood ratio (up to constants), 
reweighting the calibration data points with weights $\widehat{w}(X_i)$
ensures coverage with respect to the actual target distribution $\PP$.

Building on the framework of conformal prediction with covariate shift \citep{tibshirani2019conformal}, 
\citet{candes2021conformalized}'s result proves that this construction yields a valid LPB.
\begin{theorem}[Proposition 1 of~\citet{candes2021conformalized}]
Suppose $(X_i,C_i,T_i)\stackrel{\textnormal{iid}}{\sim} P$, and suppose 
$\widehat{w}(x) = 1/\PP(C\geq c_0|X=x)$, i.e.,
this probability was fitted exactly. 
Then $\widehat{L}_{\rm fixed-cutoff}(X)$ is a marginally calibrated LPB for $T\wedge c_0$,
and therefore also for $T$.
\end{theorem}
\noindent Moreover, \citet[Theorem 2]{candes2021conformalized} establish a double robustness result: 
if either $\widehat{w}(x)$ was fitted accurately (i.e., is a good approximation of $1/\PP(C\geq c_0 \given X=x)$
or the quantile regression was fitted accurately (i.e., $\widehat{q}_\alpha(x)$ is a good approximation of the $\alpha$-quantile
of $T\given X$), then $\widehat{L}_{\mathrm{fixed-cutoff}}$ approximately satisfies the criterion for a marginally calibrated LPB.

\section{Conformalized survival analysis with adaptive cutoffs}\label{sec:method}


\subsection{Our procedure}
As before, we first partition the data into a training set $\mathcal{I}_1$ and a calibration
set $\mathcal{I}_2$. 
On the training set, we fit a {\em family} of estimated quantile regression functions, $(x,a)\mapsto \widehat{q}_a(x)$,
mapping $x$ to the estimated $a$-quantile of the conditional distribution of $T$ given $X = x$, for all $a\in[0,1]$.
We assume that, for any $x$, $a\mapsto\widehat{q}_a(x)$ is nondecreasing.\footnote{
If our estimators $\widehat{q}_a$ are computed independently for each $a$ and this constraint
is violated, monotonicity can easily be restored via sorting the outputs---see e.g.,~\citet{koenker1994confidence}. $\widehat{q}_a$ is defined to be $-\infty$ at $a=0$.}
(In contrast, for the existing methods defined in Section~\ref{sec:background}, this regression is run only
at a single value of $a$.)

Next, we need to use the calibration set in order 
 to choose a value of $a$, for which returning $\widehat{q}_a(X)$
is a valid LPB---that is, we need to find a value $a$ such that
$ \PP\big(T < \widehat{q}_a(X)\big) = \alpha$
(where we implicitly treat the fitted quantile function $\widehat{q}_a$ as fixed,
and take the probability over $(X,T)\sim P$).
Note that, if the original regression were estimated perfectly, then we would expect to return $a=\alpha$,
i.e., the estimated $\alpha$-quantile $\widehat{q}_\alpha(X)$ would already be a valid LPB. In practice, 
as discussed earlier,
 we expect to see overfitting in most real-data settings and thus we expect to return $a < \alpha$. To choose
$a$ appropriately, we could consider solving for $a$ in the following expression:
\begin{align}
\alpha
\notag&= \PP\big(T < \widehat{q}_a(X)\big)
=\EE\left[\PP(T < \widehat{q}_a(X)\mid X)\right]\\
\notag&=\EE\left[\PP(T < \widehat{q}_a(X)\mid X) \cdot \frac{\PP(\widehat{q}_a(X) \leq C \mid X)}{\PP(\widehat{q}_a(X) \leq C \mid X)}\right]\\
\notag&\approx \EE\left[\PP(T < \widehat{q}_a(X)\mid X) \cdot \PP(\widehat{q}_a(X) \leq C \mid X) \cdot \widehat{w}_a(X)\right]\\
\notag&= \EE\left[\PP(T < \widehat{q}_a(X) \leq C \mid X) \cdot \widehat{w}_a(X)\right]\\
\label{eqn:find_a_prelim}&= \EE\left[\mathbbm{1}\{T < \widehat{q}_a(X) \leq C \} \cdot \widehat{w}_a(X)\right],
\end{align}
where now $\widehat{w}_a(x)$ is chosen to be (approximately) equal to $1/\PP(C \geq \widehat{q}_a(X)\mid X =x)$, and where
the next-to-last step holds by Assumption~\ref{assumption:conditionally_independent_censoring}.
If instead we only assume that our estimate $\hat{w}_a(x)$ is {\em proportional}
to $1/\PP(C \ge \hat{q}_a(X) \given X=x)$, then we want to solve for $a$ in the equation
\begin{align}
\alpha \notag = \PP\big(T < \widehat{q}_a(X)\big)
& \approx \frac{\EE\big[\PP(T < \hat{q}_{a}(X) \given X) \cdot 
\PP(\hat{q}_a(X) \le C \given X ) \cdot \hat{w}_a(X)\big]}
{\EE\big[\PP(\hat{q}_a(X) \le C \given X) \cdot \hat{w}_a(X) \big]}\\
& \label{eqn:find_a_prop}= \frac{\EE\big[\mathbbm{1}\{T < \hat{q}_{a}(X)  \le C \} 
\cdot \hat{w}_a(X)\big]}
{\EE\big[\mathbbm{1}\{\hat{q}_a(X) \le C \} \cdot \hat{w}_a(X) \big]}.
\end{align}
Now we can note that, while the events $\mathbbm{1}\{T_i < \widehat{q}_a(X_i) \}$ cannot be observed
on the calibration set (since we only observe the censored survival time, $\Tilde{T}_i$),
the filtered events $\mathbbm{1}\{T_i < \widehat{q}_a(X_i)\leq C_i \}$ {\em can} be observed
(since if $C_i \geq \widehat{q}_a(X_i)$, then $\mathbbm{1}\{T_i < \widehat{q}_a(X_i) \}=\mathbbm{1}\{\Tilde{T}_i < \widehat{q}_a(X_i) \}$).
Therefore, the calibration set can indeed be used to find a value $a$ so that the 
equation~\eqref{eqn:find_a_prelim} or~\eqref{eqn:find_a_prop} is (approximately) satisfied.

Now we formally describe how to select $a$ using the calibration set. For each value $a$,
we estimate the miscoverage rate $\PP(T < \widehat{q}_a(X))$ as follows:
\[\widehat{\alpha}(a) = \frac{\sum_{i\in\mathcal{I}_2} \widehat{w}_a(X_i) \cdot\mathbbm{1}\{T_i < \widehat{q}_a(X_i)\leq C_i\} 
}{\sum_{i\in\mathcal{I}_2} \widehat{w}_a(X_i) \cdot\mathbbm{1}\{ \widehat{q}_a(X_i) \leq C_i\}} .
\]
This empirical quantity estimates $\alpha^*(a) = \PP(T < \widehat{q}_a(X))$. Since our aim is to find a value of $a$
sufficiently small so that $\alpha^*(a)\leq \alpha$, we will instead search for $a$
satisfying $\widehat{\alpha}(a)\leq \alpha$.
However, while $\alpha^*(a)$ is monotone in $a$ (since $\widehat{q}_a(x)$ is monotone in $a$),
this property may not hold for the estimator $\widehat{\alpha}(a)$; we therefore
define
$\widehat{a} = \sup\big\{a\in [0,1] : \sup_{a'\leq a}\widehat{\alpha}(a')\leq \alpha\big\}$.
Finally, 
we output the LPB
$\widehat{L}(X) := \widehat{q}_{\widehat{a}}(X)$.

Below, we will give a a double robustness result
proving that this choice of $\widehat{L}$
is (approximately) a marginally calibrated LPB, as long as {\em either} the weights
$\widehat{w}_a$ {\em or} the quantiles $\widehat{q}_a$ are fitted accurately;
furthermore, when the quantiles are fitted accurately, the 
LPBs are (approximately) conditionally valid. 
Before giving our theoretical results, we first present a more general form
of this procedure.

\subsection{A generalized procedure}
\label{sec:general_procedure}
The procedure described above tends to perform well in settings where
$\widehat{q}_a(x)$ (for relevant values of $a$) is not too large---so that
$\PP(C \geq \widehat{q}_a(x) \mid X=x)$ is not close to zero and the weights
$\widehat{w}_a(X_i)$ on the calibration points are not too large.
In other settings, however, the procedure may be somewhat unstable. Specifically,
in scenarios where $C$ is often much smaller than $T$ (as in the example given
in Section~\ref{sec:intro_example} above), 
we might have a very small probability $\PP(C \geq \widehat{q}_a(X)\mid X =x)$; 
this is problematic since the inverse weight
$\widehat{w}(x)$ will then be extremely large. To alleviate this, 
we now generalize the procedure sketched above to allow for a more
stable and robust method. Define a family of functions
for $x \in \calX$ and $a\in \calA$, $(x,a)\mapsto \widehat{f}_a(x)$,
which are fitted on the training set, such that for each fixed $x$ this map is nondecreasing in $a$
(note that the estimated quantiles, $\widehat{q}_a(x)$, are simply a special case).
Our aim is now to use the calibration set in order to choose $a$ so that $\widehat{L}(X) = \widehat{f}_a(X)$
offers a calibrated LPB. With the same rationale as before, we wish to find $a$ to satisfy
\begin{align}
\alpha
\notag= \PP(T < \widehat{f}_a(X))
&\approx 
\frac{\EE\left[\PP(T < \widehat{f}_a(X)\mid X) \cdot \PP(\widehat{f}_a(X) \leq C \mid X) \cdot \widehat{w}_a(X)\right]}
{\EE\big[\PP(\hat{f}_a(X)\le C \mid X) \cdot \hat{w}_a(X) \big]}\\
\label{eqn:find_a}&= 
\frac{\EE\left[\mathbbm{1}\{T < \widehat{f}_a(X) \leq C \} \cdot \widehat{w}_a(X)\right]}
{\EE\big[\mathbbm{1} \{\hat f_a(X) \le C \} \cdot \hat{w}_a(X) \big]},
\end{align}
where $\widehat{w}_a$ is again fitted on the training data but is now chosen to be (approximately)
proportional to $1/\PP(C\geq \widehat{f}_a(X)\mid X=x)$.

From this point on, we proceed exactly as before, but with $\widehat{f}_a$ in place of $\widehat{q}_a$---we 
define
\[\widehat{\alpha}(a) = \frac{\sum_{i\in\mathcal{I}_2} \widehat{w}_a(X_i) \cdot\mathbbm{1}\{T_i < \widehat{f}_a(X_i)\leq C_i\} 
}{\sum_{i\in\mathcal{I}_2} \widehat{w}_a(X_i) \cdot\mathbbm{1}\{ \widehat{f}_a(X_i) \leq C_i\} 
},\]
which estimates $\alpha^*(a) = \PP(T < \widehat{f}_a(X))$.
As before, we compute
\begin{equation}\label{eqn:find_a_for_f}\widehat{a} = \sup\Big\{a\in [0,1] : \sup_{a'\leq a}\widehat{\alpha}(a')\leq \alpha\Big\},\end{equation}
and return the LPB $\widehat{L}(X) := \widehat{f}_{\widehat{a}}(X)$.

\textbf{Choosing the family of bounds.}
In this more general procedure, how should the family $\widehat{f}_a(x)$ be chosen?
The LPB will be approximately valid regardless of our choice, but the utility of the method
will depend strongly on choosing a reasonable family of functions.
We consider two goals when choosing the family:
\begin{itemize}
\item We would like
to closely approximate the ``oracle'' LPB, $L(X) = q_\alpha(X)$, where $q_\alpha(x)$ is the true $\alpha$-quantile
of $T$ given $X=x$. 
As a result, for $a$ such that $\hat{q}_a(x)$ is close to $q_{\alpha}(x)$, we would like to have $\widehat{f}_a(x)\approx \widehat{q}_a(x)$, our estimated
quantiles for $T$ given $X$.
\item On the other hand, we would like for the weights $\widehat{w}_a(x)$ to not be too large,
or equivalently, for $\PP(C\geq \widehat{f}_a(X) | X=x)$ to not be too small for any $a$. 
Consequently, we might want to require $\widehat{f}_a(x)\leq \widehat{q}^C_{1-\beta}(x)$,
where $\widehat{q}^C_{1-\beta}(x)$ estimates the conditional quantile of $C$ given $X=x$, and
we choose some constant value $\beta$.
\end{itemize}
To balance between these two goals, we propose selecting
$\widehat{f}_a(x) = \min\left\{ \widehat{q}_a(x), \widehat{q}^C_{1-\beta}(x)\right\}$.
As for the choice of $\beta$, we use  $\beta=1/\log|\calI_2|$ 
in our implementation such that $\hat{w}_{a}(X_i) \le \log |\calI_2| 
=o(\sqrt{|\calI_2|})$.
In the simulations, we will compare this choice against the 
``canonical'' version of the method with $\widehat{f}_a(x) = \widehat{q}_a(x)$, 
to see how this new choice adds stability to the method.

\textbf{Implementation details.}
Next we describe how the threshold $\widehat{a}$ in~\eqref{eqn:find_a_for_f} can be computed
efficiently in practice.
We note that $\sup_{a' \le a}\hat{\alpha}(a')$
is a non-decreasing piecewise constant function in $a$, 
with no more than $2n$ knots---values of $a$ at which the indicators 
$\mathbbm{1}\big\{T_i < \hat{f}_a(X_i) \le C_i\big\}$ or 
$\mathbbm{1}\big\{\hat{f}_a(X_i) \le C_i\big\}$ 
change signs. 
Denote 
$\bar{a}_i = 
\sup_{a\in[0,1]} \big\{\hat{f}_a(X_i) \leq \widetilde{T}_i\big\}, 
\tilde{a}_i = \sup_{a\in[0,1]} \big\{\hat{f}_a(X_i) \le C_i\big\}$
and 
$\calA_1 = \big\{\bar{a}_i:i=1,\dots,n\big\}, 
\calA_2=\big\{\tilde{a}_i: i=1,\dots,n\big\}.$ 
Then by definition, 
the breakpoints of the piecewise constant map $a\mapsto \widehat\alpha(a)$ must all
lie in $\calA_1\cup\calA_2$.
In the implementation, in order to obtain $\hat a$,
we only need to search through the finite grids
\begin{align}
\label{eq:grid}
\calA = \calA_1 \cup \calA_2 \cup \{0\}.
\end{align}
A complete description of the general procedure can be
found in Algorithm~\ref{alg1}.
\begin{algorithm}[htb!]
\caption{Conformalized survival analysis with adaptive cutoffs}\label{alg1}
\textbf{Input:}  Level $\alpha$; data $\calD = (X_i,\tilde{T_i},C_i)_{i\in[n]}$.
\\
\textbf{Procedure:}\\
1. Split the data into 
two folds: the training fold $\calI_1$ and 
the calibration fold $\calI_2$.\\
2. Using $\calI_1$ as input, apply any algorithm 
to fit the candidate LPBs
$\big\{\hat{f}_a(\cdot)\big\}_{a\in[0,1]}$.\\
3. Using $\calI_1$ as input, apply any algorithm 
to construct estimates
$\widehat{w}_a(x)$ of $\PP(C\geq \hat{f}_a(x)\mid X=x)$.\\
4. Determine $\calA$ according to~\eqref{eq:grid}.\\
5. \textbf{For} $a$ in $\calA$ \textbf{do:}
Compute the estimated miscoverage rate
\begin{align}
\hat{\alpha}(a) = \frac{\sum_{i \in \calI_2}
\hat{w}_{a}(X_i) \cdot  \mathbbm{1}\big\{T_i < \hat{f}_a(X_i) \le C_i\big\}}
{\sum_{i \in \calI_2} \hat{w}_{a}(X_i) \cdot
\mathbbm{1}\big\{\hat{f}_a(X_i) \le C_i\big\}}.
\end{align}
6. Compute the threshold: 
$\hat{a} = \sup\big\{a \in \calA: 
\sup_{a' \le a, a' \in \calA} 
\hat{\alpha}(a') \leq \alpha\big\}$.\\
\textbf{Return:} The calibrated LPB:
$\hat{L}(\cdot) = \hat{f}_{\hat a}(\cdot)$.
\end{algorithm}

\textbf{Computational complexity.}
The computational cost of our proposed procedure can be decomposed into  
the cost of model fitting on $\calI_1$ and that of 
finding $\hat{a}$ on $\calI_2$. The cost of the first stage 
heavily depends on the type of models chosen by the user.
For the second stage, we first need to find the set of ``knots'' $\calA$
defined in~\eqref{eq:grid}. For each $i \in \calI_2$, finding $\bar{a}_i$
(resp.~$\tilde{a}_i$) requires finding the supremum over $a$ such that $\widehat{f}_a(X_i) \le \tilde{T_i}$
(resp.~$\widehat{f}_a(X_i) \le C_i$). Since $\widehat{f}_a(X_i)$ is nondecreasing in $a$,
finding $\bar{a}_i$ or $\tilde{a}_i$ can be done efficiently via binary search.
More specifically, given a tolerance level $\epsilon$, we can obtain an $\epsilon$-accurate 
solution within $O(\log(1/\epsilon))$ runs. Repeating the above for all $i\in \calI_2$ 
requires $O(|\calI_2|\cdot \log(1/\epsilon))$ runs. Finally, 
evaluating $\hat{\alpha}(a)$ for $a\in\calA$ and finding $\hat{a}$
requires $O(|\calI_2|)$ runs. Overall, the computational complexity of 
the second stage is of the order $O(|\calI_2|\cdot(1+\log(1/\epsilon)))$.

\subsection{Theoretical guarantee: a double robustness result}

In this section, we establish the theoretical guarantees
for the LPBs produced by Algorithm~\ref{alg1}. In particular,
we show that the LPBs enjoy a double-robustness property 
in the following sense: the LPBs are approximately marginally
calibrated if {\em either} the censoring mechanism
{\em or} the conditional quantile of survival times can be estimated 
well; when the latter is true, the LPBs are furthermore 
approximately conditionally calibrated.

Given the class of functions $\{\hat{f}_a(\cdot)\}_{a\in[0,1]}$, we
define the oracle weights 
$
w_a(x) = \big(\PP\{C \ge \hat{f}_a(X) \given \calI_1, X = x\} \big)^{-1},
$
(here we condition on $\hat{f}_a$, i.e., the function is treated as fixed),
and the following oracle quantity  for any $\beta \in [0,1]$:
\$
a(\beta) = \sup\Big\{a \in [0,1]: 
\PP\big(T < \hat{f}_a(X) \given \calI_1\big)
\le \beta\Big\}.
\$
Theorem~\ref{thm:double_robustness_c} and~\ref{thm:double_robustness_t} 
develop the coverage guarantee for the LPBs.
\begin{theorem}
\label{thm:double_robustness_c}
Fix any $\delta,\alpha \in(0,1)$. Assume that
$\hat{f}_a(x)$ is continuous in $a$, and 
that for any $a\in[0,1]$, there exists 
some constant $\hat\gamma_a>0$ such that
$\hat{w}_a(x) \le \hat{\gamma}_a$ 
for $P_X$-almost all $x$. Then with probability at least $1-\delta$
over the draw of $\calD$,
the LPB produced by Algorithm~\ref{alg1} satisfies 
\$
& \PP_{(X,T)\sim P}\big(T \ge \hat{L}(X) \given \calD\big) 
\ge 1- \alpha \\
& \qquad \qquad - 
\sup_{a\in[0,1]}~\Bigg(\EE\bigg[\Big|\frac{\hat{w}_a(X)}{w_a(X)\hat{\pi}_a}-1\Big|\Biggiven \calI_1\bigg]
+\sqrt{\frac{1+\frac{\hat \gamma^2_a}{\hat{\pi}_a^2}+\max(1,\frac{\hat \gamma_a}{\hat{\pi}_a}-1)^2}{|\calI_2|}
\cdot \log\Big(\frac{1}{\delta}\Big)}\Bigg),
\$
where the probability is taken with respect to a 
new data point $(X,T)\sim P_{(X,T)}$, and where we define $\hat{\pi}_a=\EE_{X\sim P_X}\bigg[\frac{\hat{w}_a(X)}{w_a(X)}\Biggiven \calI_1\bigg]$
for any $a \in [0,1]$.
\end{theorem}
The proof of Theorem~\ref{thm:double_robustness_c}
if deferred to supplementary material.
In other words, if the estimates $\hat{w}_a$ 
are accurate approximations of $w_a$ (up to rescaling by a constant), then we have $\hat{w}_a(X)/w_a(X)\hat\pi_a\approx 1$, and
approximate coverage is guaranteed. 

Next, we show
that we also achieve approximate coverage when $T \given X$
can be accurately modeled.
\begin{theorem}
\label{thm:double_robustness_t}
Fix any $\delta,\alpha \in(0,1)$. 
Assume the same conditions as Theorem~\ref{thm:double_robustness_c},
and assume further that the conditional distribution
of $T\given X$ is continuous, with its conditional 
density upper bounded by a constant $B>0$, and that
there exists a constant $r > 0$ such that
\begin{enumerate}[(a)]
\item 
$
\sup_{\xi \in [a(\alpha),a(\alpha+r)+\psi]} 
w_{\xi}(x) \le \gamma
$
and 
$
\sup_{\xi \in [a(\alpha),a(\alpha+r)+\psi]} 
\hat{w}_{\xi}(x) \le \hat \gamma
$
for some constants $\psi,\gamma,\hat\gamma >0$;
\item 
$ 
\sup_{\beta \in [\alpha, \alpha+r]} 
\sup_{x\in \calX} \Big\{ \max(B,1) \cdot 
\big|\hat{f}_{a(\beta)}(x) - q_{\beta}(x)\big| \Big\}
+ \hat{\gamma}\gamma \sqrt{\frac{\log(1/\delta)}{|\calI_2|}} \le r,
$
where $q_{\beta}(x)$ is the $\beta$-quantile of $T$
conditional on $X = x$.
\end{enumerate}
Then with probability at least $1-\delta$ over the draw of $\calD$,
the LPB produced by 
Algorithm~\ref{alg1} satisfies that 
for $P_X$-almost all $x$,
\begin{multline*}
  \PP_{(X,T) \sim P}\big(T \ge \hat{L}(x)\given \calD, X=x\big) \ge 1 - \alpha \\ {}- 
\sup_{\beta\in[\alpha,\alpha+r]}\sup_{x\in\calX}
~\Big\{2B\cdot|\hat{f}_{a(\beta)}(x) -q_{\beta}(x)|\Big\}
- {\hat{\gamma}}{\gamma}\sqrt{\frac{1}{|\calI_2|}
\cdot \log \Big(\frac{1}{\delta}\Big)}.
\end{multline*}
\end{theorem}
The  proof of Theorem~\ref{thm:double_robustness_t}
is deferred to the supplementary material, 
where we in fact prove a more general version.
The implication of Theorem~\ref{thm:double_robustness_t}
is that if $T\given X$ can be modeled well,
the conditional miscovarege rate will be small 
(which also implies that the marginal coverage rate will be small).
\begin{remark}
\label{remark:assumption}
The assumption on the continuity $\hat{f}_a(x)$
in $a$ and the boundedness on the estimated weights
can simply be satisfied by choosing the appropriate 
class of functions in the training stage (i.e., the 
fitting procedure using $\calI_1$).
The additional assumption (a) requires
the oracle weights $w_{\beta}(x)$ is bounded
as least in a neighborhood of $a(\alpha)$; 
(b) is satisfied when $T\given X$ is estimated uniformly well in 
a neighborhood of $\alpha$ and when $|\calI_2|$ is sufficiently
large.
\end{remark}

\section{Simulations}\label{sec:sims}
We set up six synthetic experiments, and 
under each setting we generate $N=100$ i.i.d.~datasets.\footnote{The code for 
reproducing all numerical results from the simulation  and  the real data 
analysis can be found at \url{https://github.com/zhimeir/adaptive_conformal_survival_paper}.}
Each dataset consists of the training set $\calI_1$, 
calibration set $\calI_2$, and the test set $\calI_3$,
where $|\calI_1|=1,\!000$, 
$|\calI_2|=1,\!000$ and $|\calI_3|=5,\!000$. 
For all experiments, the target level is
 $1-\alpha=90\%$. In these experiments, we implement 
our proposed method with two families of bounds:
\begin{itemize}
  \item DFT-adaptive-T: the candidate LPB is given by $\hat{f}_a(x) = \hat{q}_a(x)$,
where $\hat{q}_a(x)$ is the estimated $a$-th
conditional quantile of $T$ given $X=x$.
\item DFT-adaptive-CT: the candidate LPB is given by  $\hat{f}_a(x) = 
  \min\big\{\hat{q}_a(x), \hat{q}^C_{1-\log \left(1/|\calI_2|\right)}(x)\big\}$,
where $\hat{q}_a(x)$ is as before and  
$\hat{q}^C_b (x)$ is the estimated $b$-th
conditional quantile of $C$ given $X=x$.
\end{itemize}
We also obtain LPBs
based on parametric models and other 
distribution-free methods:
\begin{itemize}
  \item Cox:  LPBs generated by the estimated Cox model 
    that is implemented as in \citet{survival-package}
  \item RandomForest: LPBs returned by
    the censored quantile regression forest \citep{li2020censored,athey2019generalized};
    the implementation is based on~\citet{li2020censored}.
    \item DFT-baseline: The distribution-free LPBs obtained by applying conformal quantile 
      regression~\citep{romano2019conformalized} to
      generating bounds for $\widetilde{T}$.
    \item DFT-fixed: conformalized LPB with a fixed thresholded 
      $c_0$; the implemental details are as suggested by~\citet{candes2021conformalized}.
\end{itemize}
For all conformalized method, the 
base algorithm for fitting 
conditional quantile of $T \given X$ is
the Cox model, and a Gaussian process 
model is fitted to approximate
$C\given X$ (this is implemented by the \texttt{GauPro}
R-package~\citep{gaupro}). 
For each dataset, we compute the following two quantities 
with the test set:
\$
\textnormal{Empirical coverage} = \frac{1}{|\calI_3|} \sum_{i\in\calI_3}
\mathbbm{1}\big\{\widehat{L}(X_i) < T_i \big\},
\qquad
\textnormal{Average LPB} = \frac{1}{|\calI_3|} \sum_{i\in\calI_3} 
\widehat{L}(X_i).
\$
An ideal method would have empirical coverage $\approx 1-\alpha$, and 
average LPB as low as possible. We shall demonstrate 
boxplots of the empirical coverage and average LPB 
resulting from the $100$ datasets.

\subsection{Synthetic setup}
We consider six data generating models, where settings 1--4 
concern univariate $X$ and settings 5--6 multivariate $X$.
For all settings, the marginal distribution 
of the covariates is given by $P_X = \text{Unif}([0,4]^p)$;
conditional on $X$, we generate $T$ and $C$ via distributions $\log T  \given  X \sim \calN\big(\mu(X),\sigma^2(X)\big)$
and $C\given X \sim P_{C \given X}$.

In settings $1$ and $2$, $p=1$ and $C \given X\sim 
\textnormal{Exp}(0.1)$---the censoring mechanism is 
completely exogenous;
in settings $3$ and $4$, $p=1$ and 
we allow $C$ to depend on $X$. 
In particular, setting 3 corresponds to the 
example shown in the introduction.
Settings 5 and 6 consider multivariate $X$, where $p=10$; 
in setting 5, $\sigma(x) = 1$, and in setting 
6 $\sigma(x)$ depends on $X$.
Table~\ref{tab:simu1}  
summarizes the parameters used in the six settings. 
\begin{table}[h]
\small
\centering 
{
\begin{tabular}{ccccc} 
Setting & $p$ & $\mu(x)$ &$\sigma(x)$ & $P_{C \given X}$ \\
$1$ & $1$ & $0.632 x$ &$2$ &\text{Exp}(0.1) \\
$2$ & $1$ & $3\cdot\mathbbm{1}\{x>2\}+x \cdot \mathbbm{1}\{x\le 2\}$ & 0.5 & \text{Exp}(0.1)\\
$3$ & $1$ & $2 \cdot \mathbbm{1}\{x>2\}+x \cdot \mathbbm{1}\{x\le2\}$ & 0.5 & \text{Exp}$\left(0.25+\frac{6+x}{100}\right)$\\
$4$ & $1$ & $3\cdot\mathbbm{1}\{x>2\}+1.5x \cdot \mathbbm{1}\{x \le2\}$
& $0.5$ & \text{lognormal}$
\big(2+  \frac{(2-x)}{50} ,0.5 \big)$\\
$5$ & $10$ & $0.126(x_1 + \sqrt{x_3x_5})+1$ & $1$ & $\text{Exp}(\frac{x_{10}}{10} + \frac{1}{20})$\\ 
$6$ & $10$ & $0.126(x_1 + \sqrt{x_3x_5})+1$ & $\frac{x_2 + 2}{4}$ & $\text{Exp}(\frac{x_{10}}{10} + \frac{1}{20})$\\
\end{tabular}
\caption{
Parameters used in the six experimental settings:\\ 
$P_X = \mathrm{Unif}([0,4]^p)$, $P_{T \given X} = \exp(\calN(\mu(X),\sigma^2(X)))$.
}
\label{tab:simu1}
}
\end{table}

\begin{figure}[ht!]
    \centering
    \includegraphics[width=\textwidth]{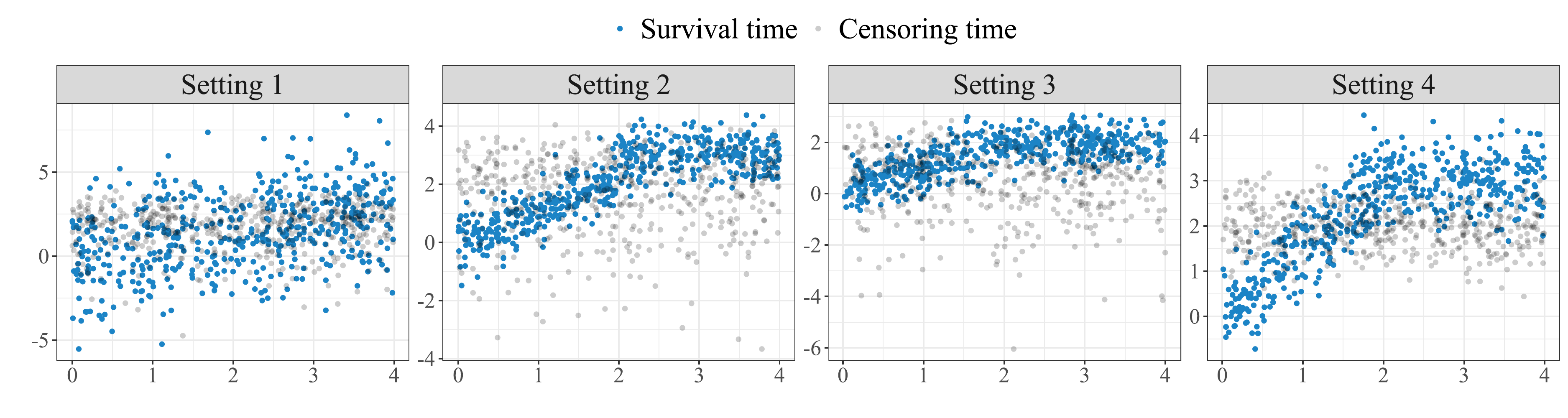}
    \caption{Illustration of the censoring time $C$ and the uncensored survival time $T$ 
    in settings 1--4 as functions of the univariate covariate $X$. 
    Both $C$ and $T$ are on a log scale.}
    \label{fig:plot_time}
\end{figure}

Figure~\ref{fig:plot_time} shows the scatterplots of the survival time $T$  
and censoring time $C$ against the univariate covariate $X$ in univariate 
experimental settings.
Settings 2--4 are more challenging than setting 1, as they all have scenarios where there are roughly two sub-populations: 
the sub-population with smaller values of $X$ has relatively higher 
censoring time, leading to a low censoring zone while the sub-population
with larger values of $X$ has comparatively higher survival time, hence 
a very high censoring zone.
In settings 5--6, there is a similar challenge: the distribution 
$P_{C\given X}$ depends on $X_{10}$ and thus we will have low censoring 
times for certain values of $X$ and higher censoring times for others.
\begin{figure}[h!]
    \centering
    \includegraphics[width=\textwidth]{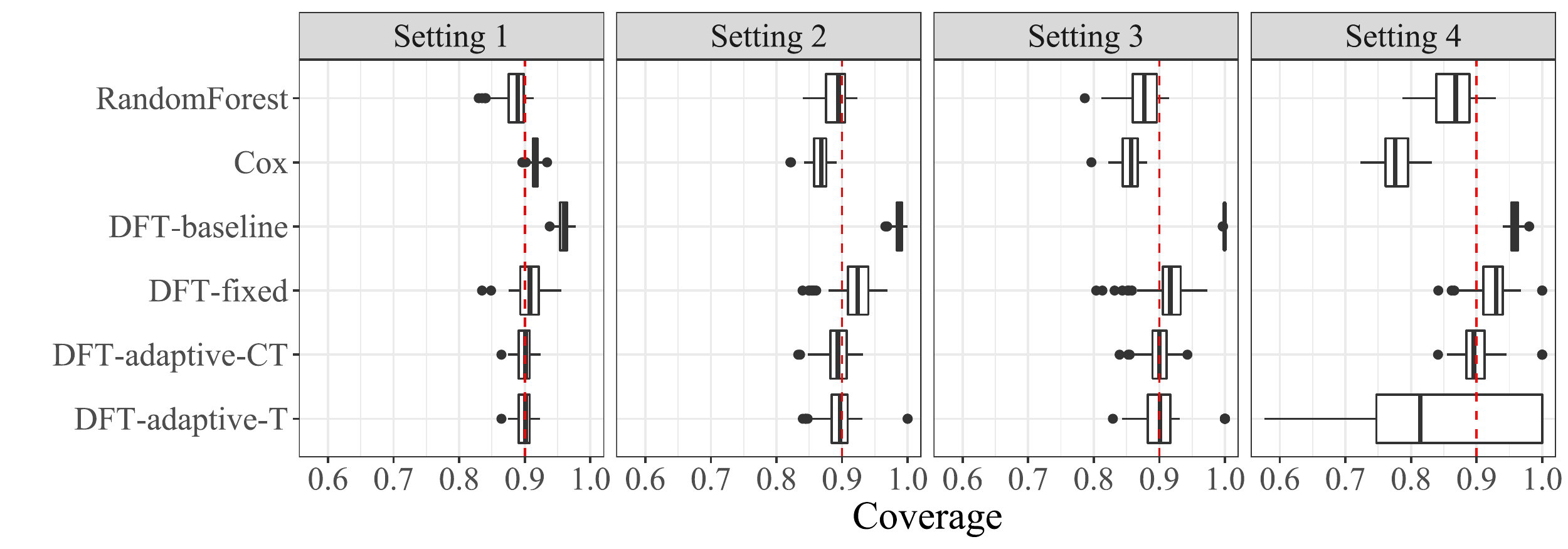}\\
    \includegraphics[width=\textwidth]{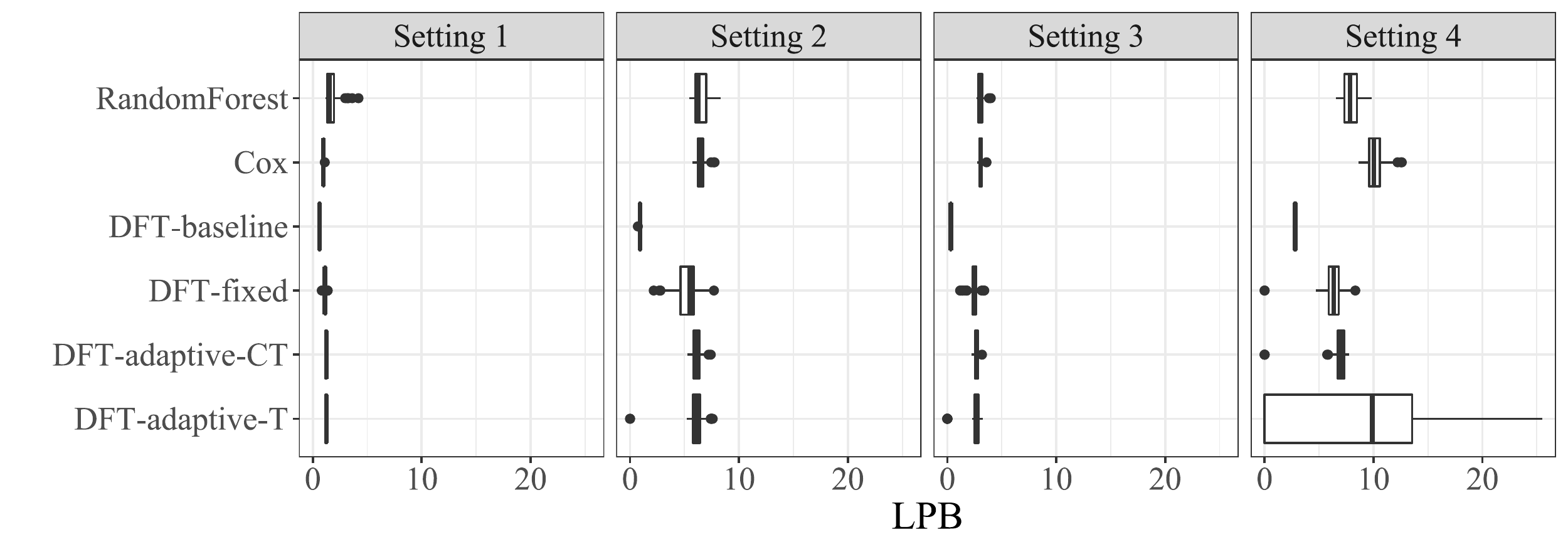}
    \caption{Empirical coverage (top) and 
        average LPBs (bottom) of all the candidate methods 
        under settings 1--4, where $X$ is univariate.
    The boxplot shows results from $100$ independent draw of datasets.
    The dashed red line corresponds to the target coverage level $1-\alpha = 90\%$.}
    \label{fig:ld}
\end{figure}

\begin{figure}[htb!]
  \begin{minipage}{0.48\textwidth}
    \centering
    \includegraphics[width=\textwidth]{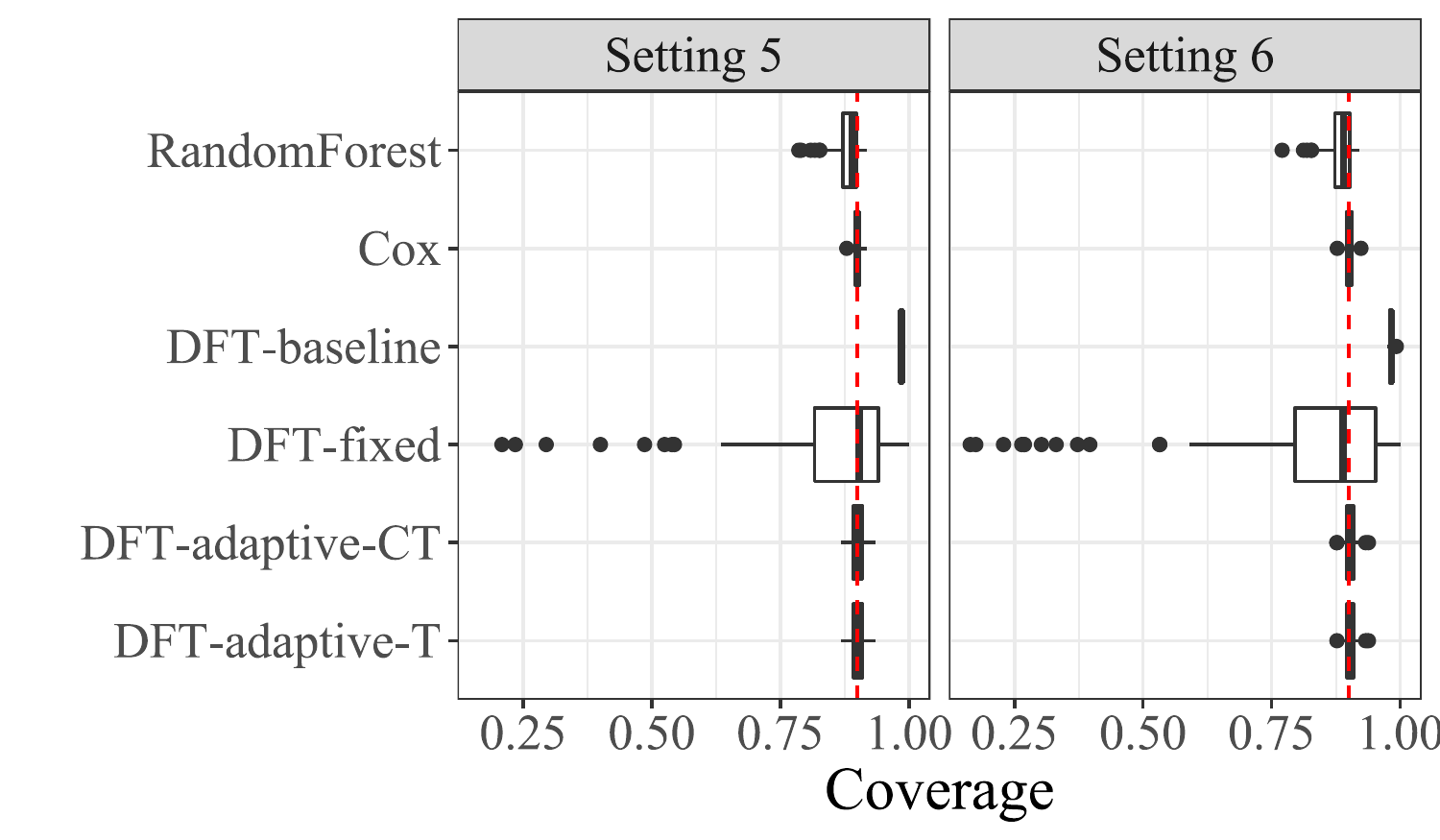}\\
  \end{minipage}
  \begin{minipage}{0.48\textwidth}
    \centering
    \includegraphics[width=\textwidth]{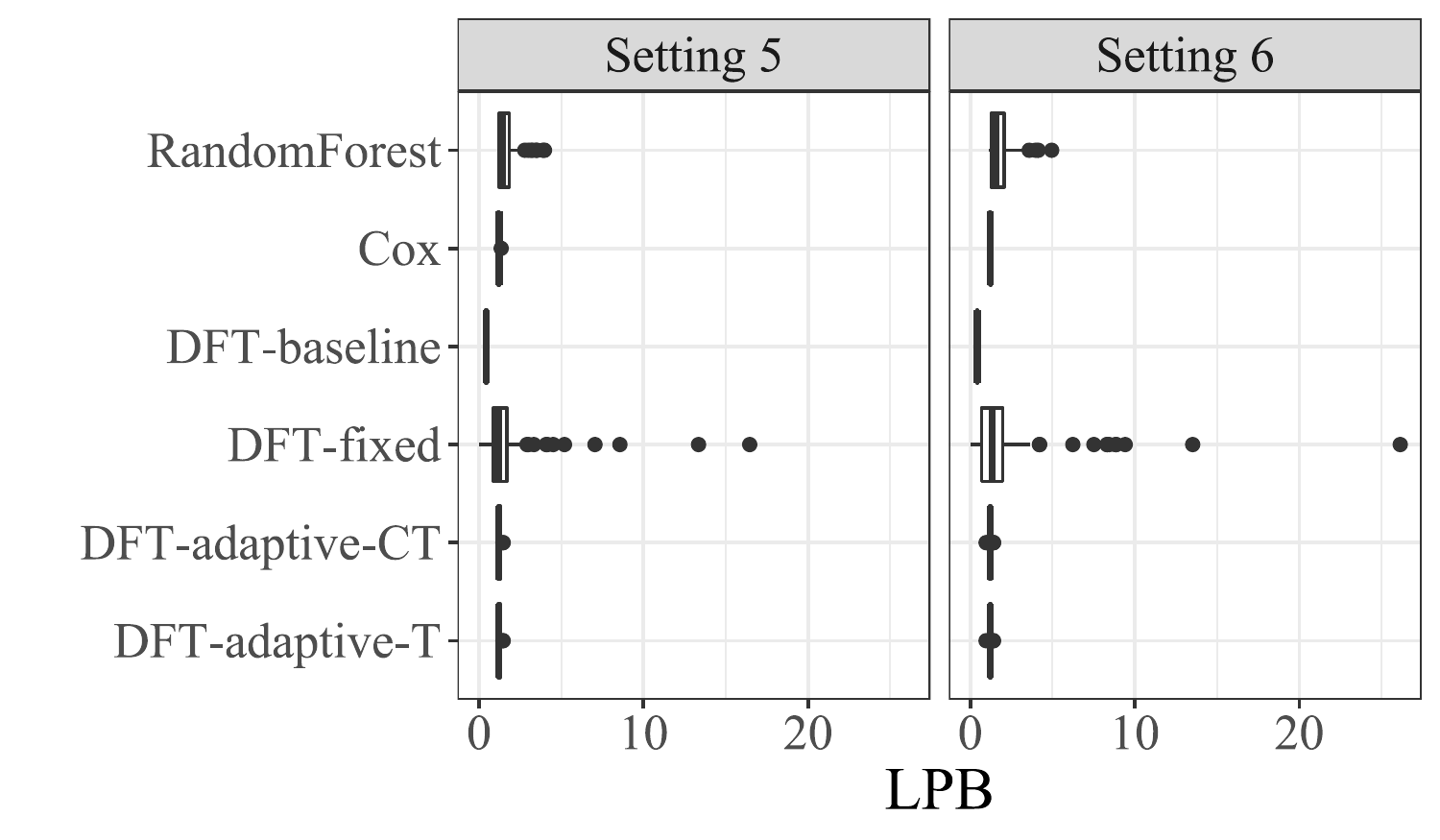}\\
  \end{minipage}
  \caption{Boxplots of empirical coverage (left) and average LPBs (right)
  in the multivariate experimental settings. The details are otherwise the 
same as in Figure~\ref{fig:ld}}
  \label{fig:hd}
\end{figure}

\subsection{Simulation results}
Figure~\ref{fig:ld} plots the empirical coverage 
and average LPBs of all candidate methods under 
the univariate settings.
First we consider the methods without distribution-free type guarantees. 
As we see in the figures, Random Forest exhibits
undercoverage in all four settings. Even in 
setting 1, a simple case where the error 
is homogeneous and $C$ does not depend on $X$, RandomForest does not return valid LPBs. Cox shows
undercoverage  as well, except for the simple regime of setting 1, and the
miscoverage gaps are even larger than those of RandomForest
in settings 2, 3,  and 4.

Next we consider the DFT methods.
The LPBs returned by  DFT-baseline are very conservative
in all four settings due to the censoring issue, 
as we expected (recall from our discussion in Section~\ref{sec:intro_censored} that this method
covers $T$ by covering the censored time $\tilde{T}$). DFT-fixed LPBs are more conservative than 
our proposed DFT-adaptive-CT LPBs, especially in settings 
2-4 where the relationship between $C$ and $T$  
changes drastically in different subpopulations. Finally, we can see
the canonical version of our method, DFT-adaptive-T, 
exhibits high variability in settings 3 and 4, verifying
our statement on stability in Section~\ref{sec:general_procedure} and highlighting the potential
advantages of DFT-adaptive-CT.

Next, Figure~\ref{fig:hd} demonstrates the results under the 
multivariate settings. Here, we observe that RandomForest 
shows slight undercoverage under both settings; Cox performs well and does not undercover, but does
not offer a distribution-free coverage guarantee. Turning to the DFT methods, DFT-baseline 
is again very conservative, and DFT-fixed LPBs exhibit high 
variability. Both of our proposed methods, DFT-adaptive-T and 
DFT-adaptive-CT are able achieve exact coverage, and the variability 
is much lower than that of DFT-fixed LPBs.

\begin{figure}[ht]
\centering
\includegraphics[width = 0.7\textwidth]{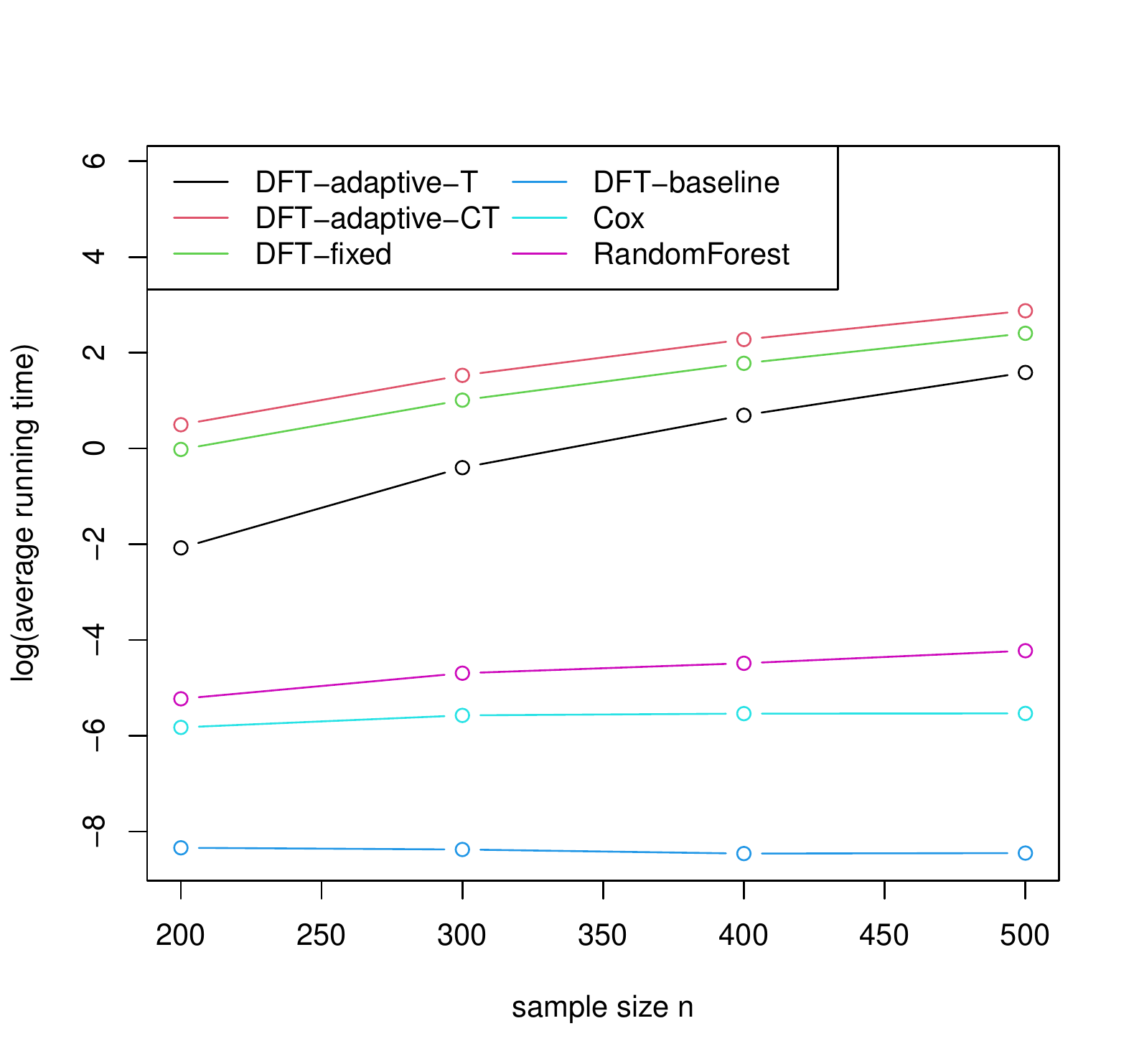}
\caption{Running time of all candidate methods in setting 3.}
\label{fig:time}
\end{figure}

Finally, we show in Figure~\ref{fig:time} the 
running time of all the candidate methods under setting 3 
(the example in Section~\ref{sec:intro_example}) with 
different choices of $n$. 
We can see that DFT-fixed, DFT-adaptive-T, and DFT-adaptive-CT 
are more computationally expensive than the other methods; the running time 
of DFT-fixed and DFT-adaptive-CT is comparable, while 
that of DFT-adaptive-T is somewhat shorter.

\section{Real data application}\label{sec:realdata}

\begin{figure}[h]
     \centering
     \begin{minipage}{0.43\textwidth}
         \centering
         \includegraphics[width=1.2\textwidth]{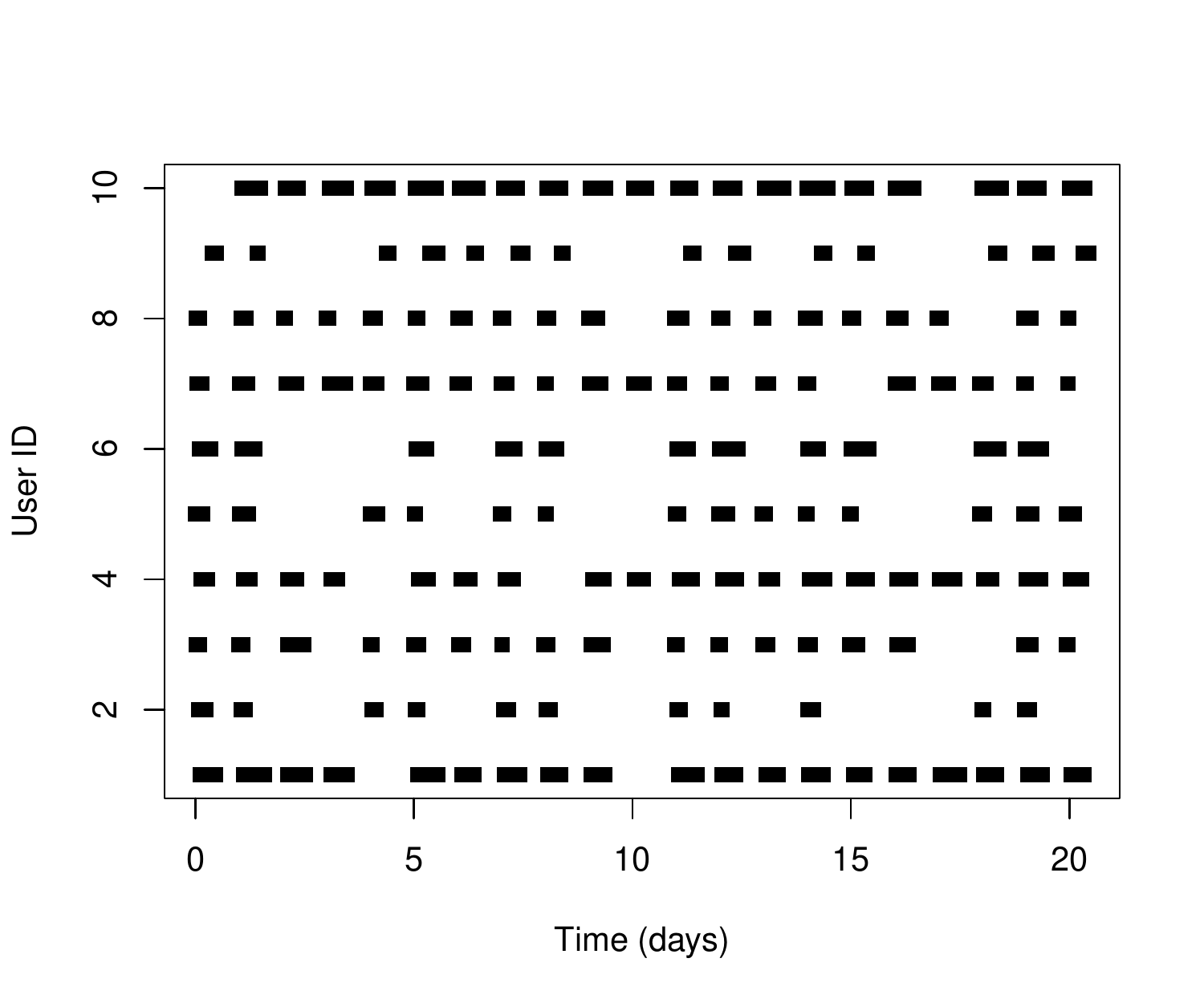}\\
         (a)
         \label{fig:plot_pingtime}
     \end{minipage}
     \hfill
     \begin{minipage}{0.43\textwidth}
         \centering
         \includegraphics[width=1.2\textwidth]{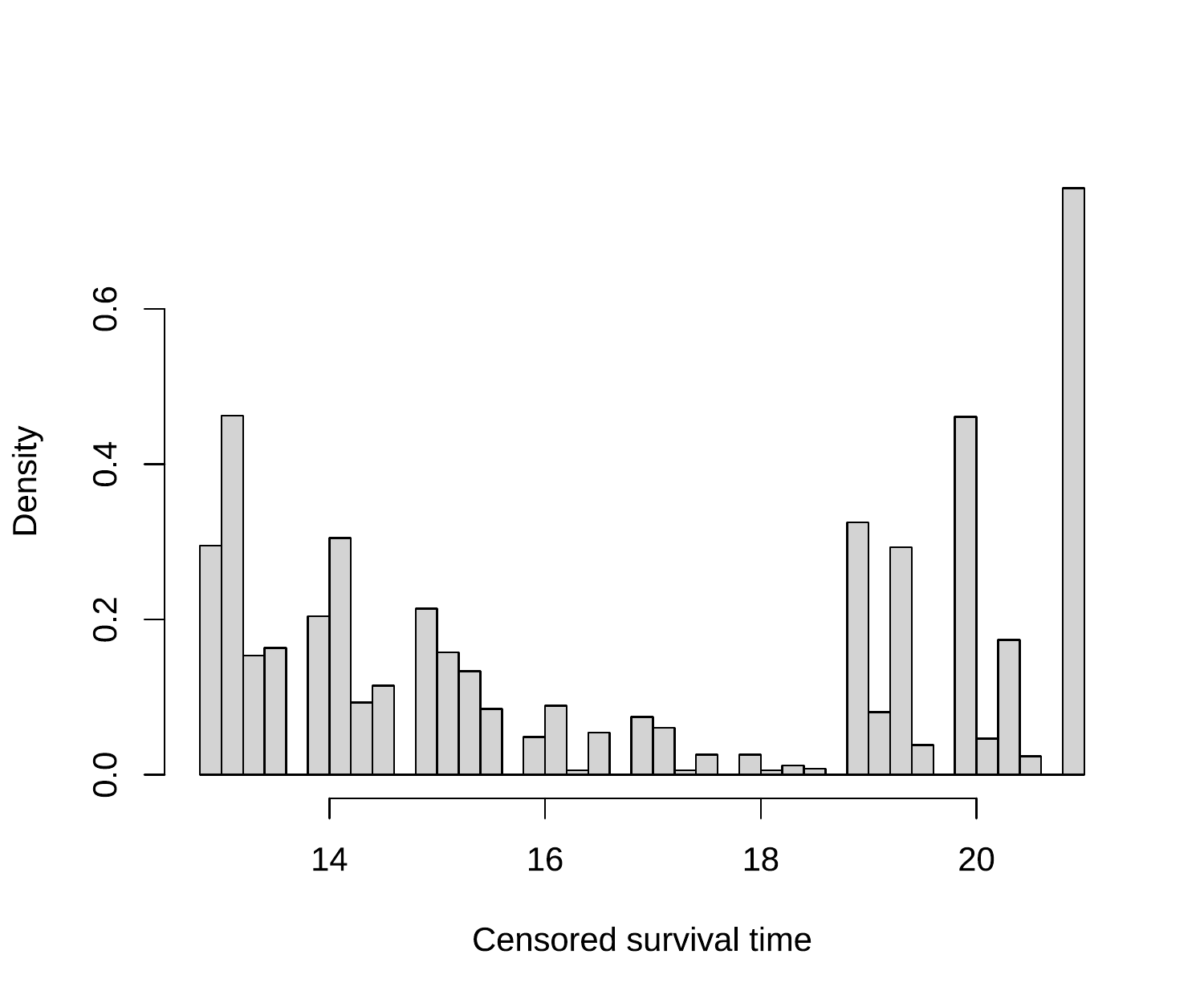}\\
         (b)
         \label{fig:hist_ping}
     \end{minipage}
     \caption{(a) Active time in three weeks for $i=1,\dots,10$; 
     (b) histogram of censored survival times for all users.  }
     \label{fig:illustration}
\end{figure}

In this section, we apply our proposed method to 
predicting users' active time on a mobile 
app with a publicly available dataset.\footnote{The data
is downloaded from
\url{https://www.kaggle.com/datasets/bhuvanchennoju/mobile-usage-time-prediction?select=pings.csv}.}
This dataset records the time stamps of 
pings for a cohort of $2,\!476$ users in 
a shared window of three weeks, where a ping 
represents a login activity or a received message. 
As is shown in Figure~\ref{fig:illustration}(a), a user's
pings gathered during an active day form a line segment,
whose length is proportional to the span of active time during that day
(the time span is standardized so that the total time window 
is mapped to the interval $[0,21]$ to represent the total number of days). 
The number of line segments and the length of line segments 
vary for different users, reflecting different types of 
user behavior. For each user, the time is recorded from the user's first active day, i.e., if the first active time for a particular
 user is $1.5$, then the sequence of active time for this user is shifted by $\lfloor 1.5 \rfloor = 1$ and is censored at time $C = 21-1=20$.

With this dataset, we focus on predicting {\it the beginning of 
a user's $14$th active day}. In practice, the prediction lower bounds can be informative if, 
for instance, the mobile app wishes to launch a promotion, offering
a discount for in-app purchases 
at the beginning of a user's $14$th active day.
For a user who is active for 
less than $14$ days within the time window, the survival time
is therefore censored and only $\widetilde{T} = \min(C,T)$ can be observed. 
Figure~\ref{fig:illustration}(b) is the histogram of the
censored survival time.
Besides the time stamps, there are three covariates in this dataset related
to users' characteristics: $X_1$ (gender), $X_2$ (age), and $X_3$ (number of children).

To implement the method, we begin by choosing
$|\calI_1|=500$ data points as the training set,
and keep this set fixed throughout.
Among the remaining data points, for 50 independent random trials,
we sample $|\calI_2| = 500$ data points as the calibration set
and another $|\calI_3| = 500$ as the test set, uniformly without replacement. 
All the methods
are applied with the target level $1-\alpha$ at $90\%$. 
Since the true survival time for censored data points are not available,
we instead empirically evaluate the upper and lower bounds of
the coverage rate:  we compute
$\beta_{\textnormal{lo}} \,:=\, \PP\big(\widetilde{T} \geq \hat{L}(X)\big) 
\leq \PP\big(T \geq \hat{L}(X)\big)$,
and also 
$\beta_{\textnormal{hi}} := 1 - \PP\big(\widetilde{T} < \hat{L}(X), T \leq C\big) \geq 
 \PP\big(T \geq \hat{L}(X)\big)$,
so that, by construction, $\beta_{\textnormal{lo}}$ is an underestimate of our target coverage rate,
and $\beta_{\textnormal{hi}}$ is an overestimate.

\begin{figure}[htb]
     \centering
     \includegraphics[width=\textwidth]{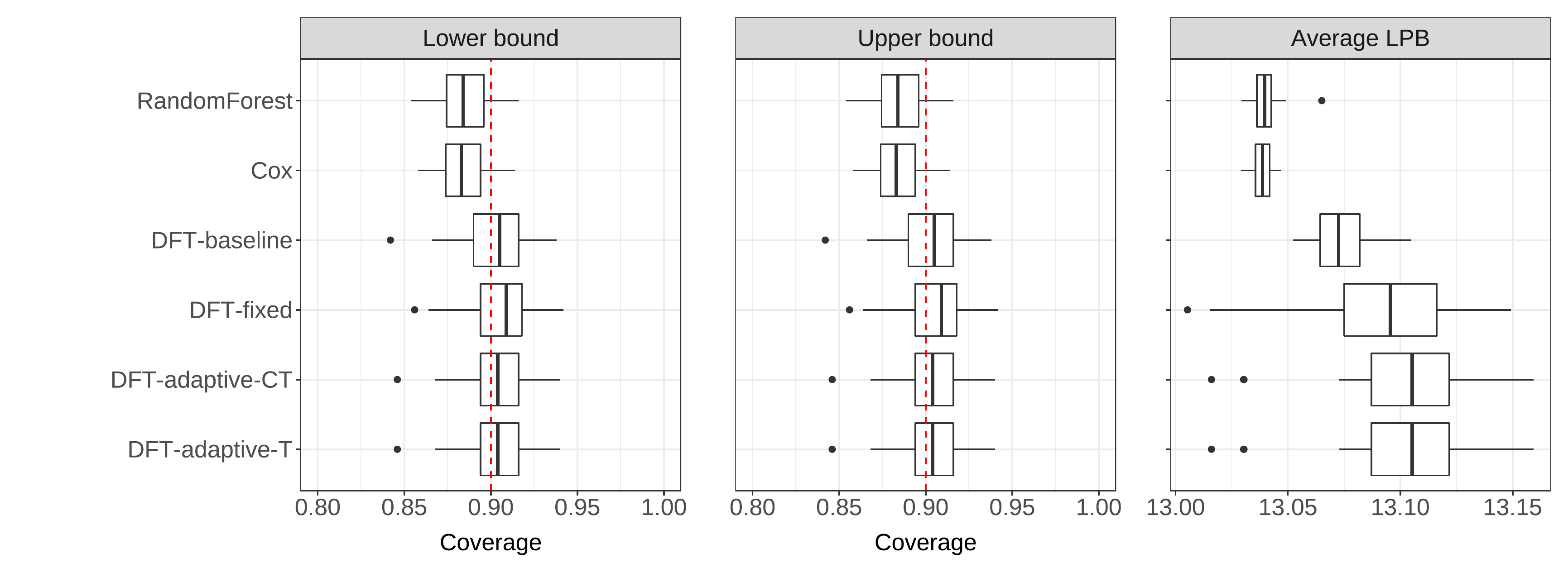}
     \caption{Left: lower bound $\beta_{\textnormal{lo}}$ of the empirical coverage rate; 
     middle: upper bound $\beta_{\textnormal{hi}}$ of the empirical coverage rate;
     right: average LPBs.}
     \label{fig:hist_ping_res}
\end{figure}

The upper and lower bounds for methods in comparison are reported in Figure~\ref{fig:hist_ping_res}.
Since this setting has a low censoring rate (19.7\%), the difference between DFT-adaptive-T
and DFT-adaptive-CT is negligible. 
Both DFT-adaptive-T and DFT-adaptive-CT attain nearly exact coverage at $90\%$, 
while Cox and Random Forests have coverage below the target level. 
In comparison, although DFT-baseline has only slightly inflated coverage,
the average LPB is lower than our methods and is thus less accurate in practice; 
in the meantime, DFT-fixed has coverage rate slightly higher than $90\%$ and shows larger variance than 
our methods with adaptive cutoffs.

\section{Discussion}\label{sec:discussion}

This paper offers a data-adaptive tool for conformalized
survival analysis. By using covariate-dependent cutoff event to subset the data,
i.e., considering data satisfying $C\geq f(X)$ for an appropriately chosen function $f$,
our method enables higher power in a broader range of scenarios, where the distribution of $C\mid X$
can vary highly with $X$ without creating overly conservative bounds, to improve on 
earlier work using a fixed cutoff, $C\geq c_0$~\citep{candes2021conformalized}.

As in~\citet{candes2021conformalized}, this work has been primarily 
focusing on the Type I censoring, where the censoring time for 
each individual is assumed observable; this is typically 
the case when the censoring time is the termination of a study. 
Another common type of censoring time is the loss-to-follow-up censoring.
When the event is death, the loss-to-follow-up censoring time is 
not observed for patients who did not survive, and our method no longer applies.
As discussed in~\citet{candes2021conformalized}, our method can however 
provide informative LPBs beyond the setting of Type I censoring: when  we have both the 
end-of-study censoring time $C_{\text{end}}$ and the 
loss-to-follow-up censoring time $C_{\text{loss}}$, the censored 
survival time is then given by $\tilde{T} = T \wedge C_{\text{end}}\wedge C_{\text{loss}}$.
Under the assumption that $(T,C_{\text{loss}}) \,\indep\, C_{\text{end}} \given X$, 
we can treat $T' := T \wedge C_{\text{loss}}$ as the true survival time and apply 
our procedure, producing an LPB on $T'$. We can thus alleviate the conservativeness 
caused by $C_{\text{end}}$, especially in studies with short duration.

We close the paper by a discussion on extensions 
and interesting directions
for future work.
First, the theoretical guarantees shown in this work focus on constructing the 
PAC-type LPB. It can also be of interest to see if 
one can derive marginal guarantees for the
proposed method, where the weighted conformal
inference technique is not applicable. (Recent work by~\citet{angelopoulos2022conformal}
on a related problem suggest tools for converting a PAC-type bound to a finite-sample
bound in expectation, and may be applicable to the survival analysis setting as well.)
Second, as with many double-robustness type results, our theoretical
guarantees rely on high accuracy of our estimate of {\em either} the 
conditional distribution of $C\mid X$ or of $T\mid X$, but 
it may be possible to establish a better bound where moderately accurate estimates
of {\em both} distributions contribute multiplicatively to a single unifying bound;
this may be more relevant to practical settings, where we might expect
moderate accuracy for each estimation problem.
Finally, as discussed earlier, the cutoff introduces a variance-bias
tradeoff---with a large cutoff, the observed survival time 
is closer to the true survival time but the effect sample 
size is reduced, and vice versa. It is interesting
to quantitatively characterize this phenomenon, and derive
an optimal choice of candidate LPBs based on this characterization.

\subsection*{Acknowledgement}
Z.R. and R.F.B were supported by the Office of Naval Research via grant N00014-20-1-2337. R.F.B. was
additionally supported by the National Science Foundation via grants DMS-1654076 and DMS-2023109.
\bibliographystyle{apalike}
\bibliography{ref}

\appendix
\section{Proofs}
\label{appx:proof}
\subsection{Proof of Theorem 3}
\label{appx:proof_double_robustness_c}
For notational convenience, we define the error term to be
\$
\Delta = \sup_{a\in[0,1]}~
\Bigg(\EE\bigg[\Big|\frac{\hat{w}_a(X)}{w_a(X)\pi_a}-1\Big|\Biggiven \calI_1\bigg]
+\sqrt{\frac{1+\frac{\hat \gamma^2_a}{\pi^2_a}+\max(1,\frac{\hat \gamma_a}{\pi_a}-1)^2}{|\calI_2|}
\cdot \log\Big(\frac{1}{\delta}\Big)} \Bigg).
\$ 
Recall that we have defined the oracle quantity
\$
a(\alpha + \Delta) 
= \sup\Big\{a \in [0,1]: \PP\big(T < \hat{f}_a(X)
\biggiven \calI_1\big) \le \alpha + \Delta\Big\}.
\$
Suppose that we can show $1-\delta\le \PP(\hat{a} \le a(\alpha+\Delta) \given \calI_1)$.
Then we have with probability at least $1-\delta$ that 
the event $\{\hat{a} \le a(\alpha + \Delta)\}$ holds and
\$
&\PP\big(T \ge \hat{f}_{\hat a}(X) \given \calD\big)\\
\ge
&\PP\big(T \ge \hat{f}_{a(\alpha+\Delta)}(X) \given \calD\big)\\
{\ge}& 1- \alpha - \Delta\\
=&1-\alpha-
\sup_{a\in[0,1]}~\Bigg\{\EE\bigg[\Big|\frac{\hat{w}_a(X)}{w_a(X)\pi_a}-1\Big|\Biggiven \calI_1\bigg]
+\sqrt{\frac{1+\frac{\hat \gamma^2_a}{\pi^2_a}+
  \max(1,\frac{\hat \gamma_a}{\pi_a}-1)^2}{|\calI_2|}
\cdot \log\Big(\frac{1}{\delta}\Big)} \Bigg\},
\$
where the first inequality is by the monotonicity of $\hat{f}_a(\cdot)$
and the second inequality
uses the left-continuity of 
$\PP\big(T<\hat{f}_a(X)\given \calD\big)$ in $a$.

The rest of the proof is devoted to establishing  
$1-\delta\le \PP(\hat{a} \le a(\alpha+\Delta) \given \calI_1)$.
Fix an arbitrary $\varepsilon >0$. 
By the definition of $\hat{\alpha}(a(\alpha + \Delta)+\eps)$,
we have 
\#\label{eq:a1_step1}
&\PP\Big(\hat{\alpha}\big(a(\alpha+\Delta) 
+ \varepsilon\big) \le \alpha \given \calI_1\big) \nonumber\\
\notag = & \PP\bigg(\sum_{i\in \calI_2}
\hat{w}_{a(\alpha+\Delta)+\varepsilon}(X_i) 
\cdot \ind\big\{T_i < \hat{f}_{a(\alpha+\Delta)+\varepsilon}(X_i) \le C_i\big\}\\
& \hspace{8em}\le \alpha \sum_{i \in \calI_2} \hat{w}_{a(\alpha+\Delta)+\varepsilon}(X_i)
\cdot \ind \big\{\hat{f}_{a(\alpha+\Delta)+\varepsilon}(X_i) \le C_i \big\}
\Biggiven \calI_1\bigg)\nonumber\\
= & \PP\bigg(\sum_{i\in \calI_2}
\hat{w}_{a(\alpha+\Delta)+\varepsilon}(X_i) 
\cdot\big(\ind\big\{T_i < \hat{f}_{a(\alpha+\Delta)+\eps}(X_i)\big\} - \alpha\big) 
\ind\big\{\hat{f}_{a(\alpha+\Delta)+\varepsilon}(X_i) \le C_i\big\} \le 0 \Biggiven \calI_1\bigg).
\#
For any $t>0$, we apply Markov's inequality and get 
\#
\eqref{eq:a1_step1}\le & 
\EE\bigg[\exp\Big(t \cdot \sum_{i\in\calI_2} \hat{w}_{a(\alpha+\Delta)+\varepsilon}(X_i) 
\big(\alpha - \ind\{T_i < \hat{f}_{a(\alpha+\Delta)+\eps}(X_i)\}\big)\\
& \hspace{8em} \times \ind \big\{\hat{f}_{a(\alpha+\Delta)+\eps}(X_i) 
\le C_i \big\}\Big) \Biggiven \calI_1\bigg]\nonumber\\
\notag = & \EE\bigg[\exp\Big(t \cdot \sum_{i \in \calI_2} \hat{w}_{a(\alpha+\Delta)+\varepsilon}(X_i) 
\cdot \big(\alpha - p_{a(\alpha+\Delta)+ \varepsilon}(X_i) 
+ p_{a(\alpha+\Delta)+\varepsilon}(X_i) \\
&\qquad\qquad\qquad\qquad- \ind\{T_i < \hat{f}_{a(\alpha+\Delta)+\varepsilon}(X_i)\}\big)
\cdot \ind \big\{\hat{f}_{a(\alpha+\Delta)+\varepsilon}(X_i) \le C_i \big\}\Big) \Biggiven \calI_1\bigg],
\#
where we define 
$p_a(x) \,:=\, 
\PP(T < \hat{f}_{a}(X) \given X = x, \calI_1)$
for any $a\in[0,1]$. 
Further conditioning on $(X_i,C_i)_{i \in \calI_2}$,
and using the fact that $C\,\indep\, T \mid X$ 
(Assumption 1),
we have
\#\label{eq:a1_step1.5}
\notag& \EE\Bigg[\exp\bigg(t \sum_{i\in\calI_2}
\hat{w}_{a(\alpha+\Delta)+\varepsilon}(X_i)
\ind\big\{\hat{f}_{a(\alpha+\Delta)}(X_i) \le C_i\big\}\nonumber\\ 
    \notag & \hspace{5em}\times \Big(p_{a(\alpha+\Delta)+\varepsilon}(X_i) 
- \ind\big\{T_i < \hat{f}_{a(\alpha+\Delta)+\varepsilon}(X_i)\big\}\Big)\bigg)
\bigggiven (X_i,C_i)_{i\in\calI_2},\calI_1\Bigg]\nonumber\\
\stackrel{\rm (a)}{\le} & \exp\Big(\frac{t^2}{8} 
\sum_{i\in\calI_2} \hat{w}_{a(\alpha+\Delta)+\varepsilon}^{2}(X_i) 
\cdot \ind\big\{\hat{f}_{a(\alpha+\Delta)+\varepsilon}(X_i) \le C_i \big\} \Big)
\nonumber\\
\stackrel{\rm (b)}{\le} & \exp\Big(\frac{|\calI_2|\cdot 
\hat \gamma^2_{a(\alpha+\Delta)+\eps} 
t^2}{8}\Big).
\#
Above, step (a) uses the $\frac{1}{4}$-sub-gaussianity of 
$p_{a(\alpha+\Delta)+\varepsilon} - 
\ind\{T_i \le \hat{f}_{a(\alpha+\Delta)+\varepsilon}(X_i)\}$;
step (b) follows from the boundedness assumption 
on the estimated weights. Combining~\eqref{eq:a1_step1}
and~\eqref{eq:a1_step1.5} leads to
\#\label{eq:a1_step2}
\notag\eqref{eq:a1_step1} \le & 
\exp\Big(\frac{|\calI_2| \cdot \hat \gamma^2_{a(\alpha+\Delta)+\eps} t^2}{8}\Big)\cdot
\EE\bigg[\exp\Big\{t \sum_{i \in \calI_2} \hat{w}_{a(\alpha+\Delta)+\varepsilon}(X_i) 
\ind \big\{\hat{f}_{a(\alpha+\Delta)+\varepsilon}(X_i) \le C_i\big\}\\ 
& \hspace{18em}\times \big(\alpha - p_{a(\alpha+\Delta)+\varepsilon}(X_i)\big)\Big\}\Biggiven \calI_1\bigg].
\#
We then condition on $(X_i)_{i\in\calI_2}$ and use the sub-gaussianity of 
$\ind\big\{\hat{f}_{a(\alpha+\Delta)+\eps}(X_i) \le C_i \big\} - 
w_{a(\alpha+\Delta)+\eps}(X_i)^{-1}$ to obtain the following bound:
\$
& \EE\bigg[\exp\Big\{t \sum_{i\in\calI_2}\hat{w}_{a(\alpha+\Delta)+\varepsilon}(X_i)
\big( \ind\{\hat{f}_{a(\alpha+\Delta)+\varepsilon }(X_i)\le C_i\}
- {w_{a(\alpha+\Delta)+\varepsilon}(X_i)^{-1}} \big)\\
& \hspace{15em}\times \big(\alpha - p_{a(\alpha+\Delta)+\varepsilon}(X_i)\big) \Big\}
\Biggiven (X_i)_{i\in\calI_2},\calI_1\bigg]\\
\le
& \exp\Big(\frac{t^2}{8}\sum_{i\in \calI_2} 
\big(\alpha - p_{a(\alpha+\Delta)+\varepsilon}(X_i)\big)^2 
\cdot \hat{w}_{a(\alpha+\Delta)+\varepsilon}(X_i)^2\Big)
\le \exp\Big(\frac{|\calI_2| \cdot  
\hat \gamma^2_{a(\alpha+\Delta)+\eps} t^2}{8}\Big),
\$
where we again use the boundedness of $\hat{w}_a(\cdot)$ in the last step.
With the above, we bound~\eqref{eq:a1_step2} as
\#\label{eq:a1_step3}
\eqref{eq:a1_step2} \le 
& \exp\Big(\frac{|\calI_2| \cdot \hat \gamma^2_{a(\alpha+\Delta)+\eps} t^2}{4}\Big)
 \EE\bigg[\exp\Big\{t \sum_{i\in\calI_2}
\frac{\hat{w}_{a(\alpha+\Delta)+\varepsilon}(X_i)}{w_{a(\alpha+\Delta)+\varepsilon}(X_i)}
\big(\alpha - p_{a(\alpha+\Delta)+\varepsilon}(X_i)\big)\Big\}\Biggiven \calI_1\bigg].
\#
Recall that we have defined for any $a \in [0,1]$ that
\$
\pi_a = \EE_{X \sim P_X}
\bigg[\frac{\hat{w}_a(X)}{w_a(X)}\Biggiven \calI_1 \bigg].
\$
We subsequently bound~\eqref{eq:a1_step3} as
\#\label{eq:a1_step3.5}
\eqref{eq:a1_step3}
\le & \exp\Big(\frac{|\calI_2|\cdot \hat 
\gamma_{a(\alpha+\Delta)+\eps}^2 t^2}{4}\Big)\cdot
\EE\Bigg[\exp\bigg\{t \sum_{i\in\calI_2}\pi_{a(\alpha+\Delta)+\varepsilon}\big(\alpha - p_{a(\alpha+\Delta)+\varepsilon}(X_i)\big)\nonumber\\
&  \qquad \qquad \qquad  
+ t \sum_{i\in\calI_2}\Big|\frac{\hat{w}_{a(\alpha+\Delta)+\varepsilon}(X_i)}{w_{a(\alpha+\Delta)+\varepsilon}(X_i)} - \pi_{a(\alpha+\Delta)+\varepsilon}\Big| 
\bigg\}\Biggiven \calI_1\Bigg]\nonumber\\
    \le &\exp\Big(\frac{|\calI_2| \cdot\hat \gamma_{a(\alpha+\Delta)+\eps}^2 t^2}{4}\Big)\cdot
\EE\bigg[\exp\Big\{2t \cdot \pi_{a(\alpha+\Delta)+\varepsilon} \sum_{i\in\calI_2}
\big(\alpha - p_{a(\alpha+\Delta)+\varepsilon}(X_i)\big)\Big\} \Biggiven \calI_1\bigg]^{1/2}\nonumber\\
& \qquad \qquad \qquad  \times \EE\bigg[\exp\bigg\{2 t\sum_{i\in\calI_2} 
\Big|\frac{\hat{w}_{a(\alpha+\Delta)+\varepsilon}(X_i)}
{w_{a(\alpha+\Delta)+\varepsilon}(X_i)} - \pi_{a(\alpha+\Delta)+\varepsilon}\Big|\bigg\} \Biggiven \calI_1 \bigg]^{1/2},
\#
where the last step follows from the Cauchy-Schwarz inequality.
By the definition of $a(\alpha+\Delta)$, it holds that 
\$
\PP\big(T < \hat{f}_{a(\alpha+\Delta) + \varepsilon}(X) \given \calI_1\big) \ge \alpha+\Delta. 
\$
Using the above inequality, we have
\#\label{eq:a1_31}
& \EE\bigg[\exp\Big\{2 t \cdot \pi_{a(\alpha+\Delta)+\varepsilon} \sum_{i\in\calI_2}
\big(\alpha - p_{a(\alpha+\Delta)+\varepsilon}(X_i) \big)\Big\}\Biggiven \calI_1\bigg]\nonumber\\
\le & \EE\bigg[\exp\Big\{2t \cdot \pi_{a(\alpha+\Delta)+\varepsilon} \sum_{i\in\calI_2} 
 \big(\PP(T < \hat{f}_{a(\alpha+\Delta)+\varepsilon}(X)\given \calI_1)
- \Delta - p_{a(\alpha+\Delta)+\varepsilon}(X_i) \big)\Big\} \Biggiven \calI_1\bigg]\nonumber\\
= & \exp\big(-2t\pi_{a(\alpha+\Delta)+\eps}\Delta|\calI_2|\big) \nonumber\\
&  \qquad \times \EE\bigg[\exp\Big\{2t \cdot \pi_{a(\alpha+\Delta)+\varepsilon} \sum_{i\in\calI_2} 
 \big(\PP(T < \hat{f}_{a(\alpha+\Delta)+\varepsilon}(X)\given \calI_1)
 - p_{a(\alpha+\Delta)+\varepsilon}(X_i) \big)\Big\} \Biggiven \calI_1\bigg]\nonumber\\
\le & \exp\Big(\frac{|\calI_2|\cdot t^2 \pi^2_{a(\alpha+\Delta)+\varepsilon}}{2} 
- 2 t \cdot \pi_{a(\alpha+\Delta)+\varepsilon} \Delta\cdot |\calI_2| \Big).
\#
Above, the last inequality uses that $\PP(T < \hat{f}_{\alpha(a+\Delta)+\eps} \given \calI_1)
- p_{\alpha(a + \Delta) + \eps}(X_i)$ 
is $\frac{1}{4}$-sub-gaussian.
Note that for any $i \in \calI_2$, 
\$
& \bigg|\frac{\hat{w}_{a(\alpha+\Delta)+\varepsilon}(X_i)}
{w_{a(\alpha+\Delta)+\varepsilon}(X_i)}-\pi_{a(\alpha+\Delta)+\varepsilon}(\calI_1)\bigg|\nonumber\\
\le & \tilde{\gamma}(a(\alpha+\Delta)+\varepsilon) \defn 
\max\big\{\hat \gamma_{a(\alpha+\Delta)+\eps}-\pi_{a(\alpha+\Delta)+\varepsilon},\pi_{a(\alpha+\Delta)+\varepsilon}\big\}.
\$
We then have
\begin{align}\label{eq:a1_32}
& \EE\Bigg[\exp\bigg\{2t \Big(\sum_{i\in\calI_2}
\Big|\frac{\hat{w}_{a(\alpha+\Delta)+\varepsilon}(X_i)}
{w_{a(\alpha+\Delta)+\varepsilon}(X_i)}-\pi_{a(\alpha+\Delta)+\varepsilon}\Big| \nonumber \\
& \qquad \qquad \qquad \qquad  -
\EE\Big[\Big|\frac{\hat{w}_{a(\alpha+\Delta)+\varepsilon}(X_i)}
{w_{a(\alpha+\Delta)+\varepsilon}(X_i)}-\pi_{a(\alpha+\Delta)+\varepsilon}\Big| 
\Biggiven \calI_1\Big]\Big)\bigg\} \Bigggiven \calI_1\Bigg] \nonumber\\
  \le \, &\exp\Big(\frac{|\calI_2|\cdot \tilde{\gamma}^2(a(\alpha + \Delta) + \eps) t^2}{2}\Big). 
\end{align}
Combining~\eqref{eq:a1_31} and~\eqref{eq:a1_32}, 
we have the following upper bound on \eqref{eq:a1_step3.5}:
\#
\label{eq:a1_step4}
\notag \eqref{eq:a1_step3.5}
& \le \exp\Bigg\{\frac{|\calI_2| t^2}{4} \cdot \Big({\pi^2_{a(\alpha+\Delta)+\varepsilon} 
+ \hat \gamma^2_{a(\alpha+\Delta)+\eps} 
+\tilde{\gamma}^2(a(\alpha+\Delta)+\varepsilon)}\Big) \\
&   + |\calI_2|t \cdot 
\bigg( \EE\Big[\Big|\frac{\hat{w}_{a(\alpha+\Delta)+\varepsilon}(X)}{w_{a(\alpha+\Delta)+\varepsilon}(X)} - \pi_{a(\alpha+\Delta)+\varepsilon} \Big|\Biggiven \calI_1\Big] 
- \pi_{a(\alpha+\Delta)+\varepsilon} \Delta \bigg)
\Bigg\}.
\#
We now take
\$
t = \frac{2\big(\Delta \cdot \pi_{a(\alpha+\Delta)+\varepsilon} - 
\EE\big[|\frac{\hat{w}_{a(\alpha+\Delta)+\varepsilon}(X)}{w_{a(\alpha+\Delta)+
\varepsilon}(X)} - \pi_{a(\alpha+\Delta)+\varepsilon} |
\biggiven \calI_1\big]\big)}
{\pi^2_{a(\alpha+\Delta)+\varepsilon}
+\hat \gamma^2_{a(\alpha+\Delta)+\eps} 
+ \tilde{\gamma}^2(a(\alpha+\Delta)+\varepsilon)}.
\$ 
This gives us
\begin{align*}
\eqref{eq:a1_step4} \le 
\exp\bigg(- \frac{|\calI_2| \cdot \Big(\Delta\cdot \pi_{a(\alpha+\Delta)+\varepsilon}(\calI_1)- 
\EE\big[|\frac{w_{a(\alpha+\Delta)+\varepsilon}(X)}
{w_{a(\alpha+\Delta)+\varepsilon}(X)}- \pi_{a(\alpha+\Delta)+\varepsilon}|\biggiven \calI_1\big]\Big)^2}
{ \pi^2_{a(\alpha+\Delta)+\varepsilon}(\calI_1) + \hat \gamma^2_{a(\alpha+\Delta)+\eps} 
+ \tilde{\gamma}^2(a(\alpha+\Delta)+\varepsilon)}\bigg) \le  \delta.
\end{align*}
The last inequality is due to the choice of $\Delta$.
As a result,
\$
1-\delta \le 
\PP\Big(\hat{\alpha}\big(a(\alpha+\Delta)+\varepsilon\big) > \alpha \biggiven \calI_1\Big)
\le 
\PP\Big(\hat{a} < a(\alpha+\Delta) + \varepsilon \biggiven \calI_1\Big).
\$
Since the above holds for any $\varepsilon >0$, 
we can take $\varepsilon \rightarrow 0$ and
by the continuity of the probability measure, 
we have $1-\delta\le \PP(\hat{a} \le a(\alpha+\Delta) \given \calI_1)$
and thus complete the proof.

\subsection{Proof of Theorem 4}
\label{appx:proof_double_robustness_t}
Instead of proving Theorem 4
directly, we prove a more general theorem that implies
Theorem 4.
\begin{theorem}
\label{thm:double_robustness_general}
Fix any $\delta,\alpha \in(0,1)$. 
Under the same condition of Theorem 3,
assume further that there exists a constant $r > 0$ such that
\begin{enumerate}[(a)]
  \item $\sup_{\xi \in [a(\alpha),a(\alpha+r)+\psi]}
w_{\xi}(x) \le \gamma$ and 
$\sup_{\xi \in [a(\alpha), 
a(\alpha+r)+\psi]}\hat{w}_{\xi}(x) \le \hat{\gamma}$,
for some constants $\gamma,\hat{\gamma},\psi > 0$; 
\item\label{itm:additional_asst_b}  
for any $\eta \in [0,r]$, for 
any $\beta \in [\alpha,\alpha+r]$, and $P_X$-almost all $x$,
\$
& \PP\big(T < q_{\beta}(X) + \eta \given X = x\big) \le \beta + B\eta,\\
&\PP\big(T < q_{\beta}(X) - \eta \given X = x\big) \ge \beta - B\eta,
\$
for some family of oracle functions
$\{q_{a}(\cdot)\}_{a\in[0,1]}$
and  some constant $B>0$;
\item \label{asst:c}
$ 
\sup_{\beta \in [\alpha, \alpha+r]} 
\sup_{x\in \calX} \Big\{ \max(B,1) \cdot 
\big|\hat{f}_{a(\beta)}(x) - q_{\beta}(x)\big|
\Big\}
+ \hat{\gamma}
\gamma \sqrt{\frac{\log(1/\delta)}{|\calI_2|}}\le r.
$
\end{enumerate}
Then with probability at least $1-\delta$ over the draw
of $\calD$, the LPB produced by 
Algorithm 1 satisfies that 
for $P_X$-almost all $x$,
\begin{align}\label{eq:err_t2}
&\PP_{(X,T)\sim P}\big(T \ge \hat{L}(x)\given \calD, X=x\big)\\
\ge &  1 - \alpha - 
\sup_{\beta\in[\alpha,\alpha+r]}\sup_{x\in\calX}
~\Big\{2B\cdot|\hat{f}_{a(\beta)}(x) -q_{\beta}(x)|
\Big\}
  - \hat{\gamma} \gamma
  \sqrt{\frac{1}{|\calI_2|}
\cdot \log \Big(\frac{1}{\delta}\Big)}.
\end{align}
\end{theorem}
To see why Theorem~\ref{thm:double_robustness_general}
implies Theorem 4, note that
when we take $q_{\beta}(x)$ to be the $\beta$-quantile
of $T$ conditional on $X=x$, and when the conditional 
distribution of $T\given X$ is continuous with
conditional density bounded by $B$, assumption~\ref{itm:additional_asst_b} required by
Theorem~\ref{thm:double_robustness_general} is satisfied.
We now proceed to prove Theorem~\ref{thm:double_robustness_general}.

\begin{proof}
Here, for notational convenience we define
\$
\calE = \sup_{\beta \in [\alpha, \alpha+r]}
\sup_{x\in\calX}~\big|\hat{f}_{a(\beta)}(x) - q_\beta(x)\big|
\mbox{ and }
\Delta = B \calE + \hat{\gamma}\gamma \cdot
\sqrt{\frac{\log (1/\delta)}{|\calI_2|}}.
\$
By Assumption~\ref{asst:c}, $\calE \le r$ and $\Delta \le r$. 
If we can show w.p.~at least $1-\delta$ that
$a(\alpha+\Delta)\ge \hat{a}$, then 
by the monotonicity of
$\hat{f}_a(\cdot)$ in $a$, 
\$
\PP\big(T < \hat{f}_{\hat a}(X) \given \calD\big)
\le
\PP\big(T < \hat{f}_{a(\alpha + \Delta)}(X) \given \calD\big)
\le \alpha + \Delta,
\$
where the last inequality uses the left-continuity of $\PP(T < \hat{f}_a(X)\given \calD)$.
Furthermore, for a given $x\in\calX$,  
\$
&\PP\big(T < \hat{f}_{\hat a}(x) \given X = x, \calD\big)
\le  \PP\big(T < \hat{f}_{a(\alpha+\Delta)}(x) 
\given X = x, \calD\big)\\
\le &
\PP\big(T < q_{\alpha+\Delta}(x) + \calE 
\given X = x, \calI_2\big)
\le  \alpha+\Delta + B\calE,
\$
where the second inequality follows from the definition of 
$\calE$ (and that $\Delta \le r$), and the last inequality is due to 
Assumption~\ref{itm:additional_asst_b} (and that both $\Delta$ and 
$\calE$ are bounded by $r$). We have therefore 
arrived at our desired conclusion.

It remains to show that $\PP(a(\alpha+\Delta) \ge \hat{a}) \ge 1-\delta$. 
As before, we fix an arbitrary $\varepsilon \in (0,\psi]$.
Then for any $t > 0$,
\begin{align}\label{eq:eq21}
&\PP\Big(\hat{\alpha}
\big(a(\alpha+\Delta)+\varepsilon \big) \le \alpha \biggiven \calI_1\Big) \nonumber \\
\notag = & \PP\Big(\alpha \sum_{i\in\calI_2} \hat{w}_{a(\alpha+\Delta)+\eps}(X_i) \cdot 
\ind \big\{C_i \ge \hat{f}_{a(\alpha+\Delta)+\eps}(X_i) \big\}\\
& \hspace{6em} - \sum_{i\in\calI_2} \hat{w}_{a(\alpha+\Delta)+\eps}(X_i) \cdot
\ind \big\{C_i \ge \hat{f}_{a(\alpha+\Delta)+\eps}(X_i) > T_i \big\} 
\ge 0 \biggiven \calI_1\Big)\nonumber\\
\le & \EE\Bigg[\exp\bigg\{t\cdot \Big(\alpha 
\sum_{i\in\calI_2} \hat{w}_{a(\alpha+\Delta)+\eps}(X_i) \cdot 
\ind \{C_i \ge \hat{f}_{a(\alpha+\Delta)+\eps}(X_i)\} \nonumber\\
&\qquad \qquad \qquad
- \sum_{i \in \calI_2} \hat{w}_{a(\alpha+\Delta)+\eps}(X_i)
\cdot \ind \{C_i \ge \hat{f}_{a(\alpha+\Delta)+\eps}(X_i) > T_i\} \Big)\bigg\}
\bigggiven \calI_1\Bigg],
\end{align}
where the inequality follows from Markov's inequality.
Next, we condition on $(X_i,T_i)_{i\in \calI_2}$ and have
\begin{align}\label{eq:eq21.5}
& \EE\bigg[\exp\Big\{t\sum^n_{i \in \calI_2} \hat{w}_{(\alpha + \Delta)+\eps}(X_i)
\ind\big\{C_i \ge \hat{f}_{a(\alpha+\Delta)+\eps}(X_i)\big\}
\cdot\big(\alpha - \ind\big\{T_i < \hat{f}_{a(\alpha+\Delta)+\eps}(X_i)\big\}\big)\nonumber\\
& \hspace{8em}
- t\sum_{i \in \calI_2}\frac{\hat{w}_{a(\alpha+\Delta)+\eps}(X_i)}{w_{a(\alpha+\Delta)+\eps}(X_i)}
\cdot\big(\alpha - \ind\big\{T_i < \hat{f}_{a(\alpha+\Delta)+\eps}(X_i)\big\}\big)
\Big\}\Biggiven (X_i,T_i)_{i\in \calI_2}, \calI_1\bigg]   \nonumber \\
\stackrel{\rm (a)}{\le} & \exp\Big\{\frac{t^2}{8}\sum_{i \in \calI_2}\hat{w}^2_{a(\alpha+\Delta)+\eps}(X_i)
\big(\alpha - \ind\{T_i < \hat{f}_{a(\alpha+\Delta)+\eps}(X_i)\}\big)^2 \Big\} \nonumber \\
    \stackrel{\rm (b)}{\le} & \exp\Big(\frac{|\calI_2|\cdot \hat{\gamma}^2 t^2}{8}\Big),
\end{align}
where step (a) is due to the $\frac{1}{4}$-sub-Gaussianity of 
$\ind\{C_i \ge \hat{f}_{a(\alpha+\Delta)+\eps}(X_i)\}$,
and step (b) the boundedness of $\hat{w}_{a(\alpha+\Delta)+\eps}(\cdot)$.
Combining~\eqref{eq:eq21} and~\eqref{eq:eq21.5}, we have
\begin{align}\label{eq:eq22}
\eqref{eq:eq21} \le &
\EE\Bigg[\exp\bigg\{t\sum_{i \in \calI_2}
\frac{\hat{w}_{a(\alpha + \Delta)+\eps}(X_i)}{w_{a(\alpha+\Delta)+\eps}(X_i)}
\Big(\alpha - \ind\big\{T_i < \hat{f}_{a(\alpha + \Delta)+\eps}(X_i)\big\}\Big) \nonumber\\
                    & \hspace{15em} + \frac{|\calI_2|\hat{\gamma}^2t^2}{8}\bigg\} \bigggiven \calI_1\Bigg].
\end{align}
Next, recall that 
$p_a(x) = \PP\big(T < \hat{f}_a(x) \given X=x\big)$.
We now condition on $(X_i)_{i\in\calI_2}$:
\begin{align*}
&\EE\bigg[\exp\Big\{t\sum_{i\in\calI_2}
\frac{\hat{w}_{a(\alpha+\Delta)+\eps}(X_i)}{w_{a(\alpha+\Delta)+\eps}(X_i)}
\cdot 
\big(p_{a(\alpha+\Delta)+\eps}(X_i) - 
\ind\{T_i < \hat{f}_{a(\alpha+\Delta)+\eps}(X_i)\} \big)\Big\} \Biggiven (X_i)_{i \in\calI_2},\calI_1\bigg]\\
\stackrel{\rm (a)}{\le} & \exp\Big\{\frac{t^2}{8}\sum_{i\in\calI_2}
 \frac{\hat{w}_{a(\alpha+\Delta)}(X_i)^2}{w_{a(\alpha+\Delta)}(X_i)^2}\Big\} \\
  \stackrel{\rm (b)}{\le} &\exp\Big(\frac{|\calI_2| \cdot \hat{\gamma}^2 t^2}{8}\Big),
\end{align*}
where step (a) uses the sub-Gaussianity of 
$p_{a}(X_i) - \ind\{T_i < \hat{f}_{a}(X_i)\}$
and step (b) is due to the boundedness of $\hat{w}_{a}(\cdot)$.
Combining the above, we now have
\begin{align}\label{eq:eq23}
\eqref{eq:eq22} 
\le & \EE\bigg[\exp\Big(t\sum_{i \in \calI_2} 
\frac{\hat{w}_{a(\alpha+\Delta)+\eps}(X_i)}
{w_{a(\alpha+\Delta)+\eps}(X_i)} \cdot
\big(\alpha - p_{a(\alpha+\Delta)+\eps}(X_i)\big)
+ \frac{|\calI_2|\hat{\gamma}^2 t^2}{4}\Big)\Biggiven \calI_1\bigg]\nonumber\\
= & \EE\bigg[\exp\Big(t\sum_{i\in \calI_2} 
\frac{\hat{w}_{a(\alpha+\Delta)+\eps}(X_i)}
{w_{a(\alpha+\Delta)+\eps}(X_i)} \cdot
\big(\alpha - \PP(T_i < \hat{f}_{a(\alpha+\Delta)+\eps}(X_i) \given X_i,\calI_1)\big)
+ \frac{|\calI_2|\hat{\gamma}^2 t^2}{4}\Big) \Biggiven \calI_1\bigg].
\end{align}
By Assumption~\ref{asst:c}, $\Delta \le r$, and 
by the definition of $\calE$,
$\big|q_{\alpha+\Delta}(x) - 
\hat{f}_{a(\alpha+\Delta)}(x)\big| \le \calE$;
Consequently, 
\$ 
\PP(T_i < \hat{f}_{a(\alpha+\Delta)+\eps}(X_i) \given X_i,\calI_1)
\ge  
\PP(T_i < \hat{f}_{a(\alpha+\Delta)}(X_i) \given X_i,\calI_1)
\ge 
\PP(T_i < q_{\alpha+\Delta}(X_i) - \calE \given X_i,\calI_1).
\$
Then, 
\begin{align}
\eqref{eq:eq23}
\le &
\EE\bigg[\exp\Big(t\sum_{i \in \calI_2} 
\frac{\hat{w}_{a(\alpha+\Delta)+\eps}(X_i)}
{w_{a(\alpha+\Delta)+\eps}(X_i)}\cdot
\big(\alpha - \PP(T_i < q_{\alpha+\Delta}(X_i) - \calE \given X_i, \calI_1)\big) 
+ \frac{|\calI_2|\hat{\gamma}^2 t^2}{4}\Big) \Biggiven \calI_1\bigg]\nonumber\\
\stackrel{\rm (a)}{\le} &\EE\bigg[\exp\Big(t\sum_{i\in \calI_2} 
\frac{\hat{w}_{a(\alpha+\Delta)+\eps}(X_i)}{w_{a(\alpha+\Delta)+\eps}(X_i)}\cdot
\big(\alpha - (\alpha+\Delta - B\calE) \big) 
+ \frac{|\calI_2| \hat{\gamma}^2 t^2}{4}\Big) \Biggiven \calI_1\bigg]\nonumber\\
{=} &\EE\bigg[\exp\Big(-{t(\Delta - B\calE)}\cdot\sum_{i \in \calI_2} 
\frac{\hat{w}_{a(\alpha+\Delta)+\eps}(X_i)}
{w_{a(\alpha+\Delta)+\eps}(X_i)} 
+ \frac{|\calI_2| \hat{\gamma}^2 t^2}{4}\Big) \Biggiven \calI_1\bigg]\\
 \stackrel{\rm (b)}{\le} &\exp\Big(- \frac{|\calI_2| t(\Delta - B\calE)}{\gamma}
+ \frac{|\calI_2|\hat{\gamma}^2 t^2}{4}\Big),
\end{align}
where step (a) is due to that $\calE \le r$ and Assumption~\ref{itm:additional_asst_b};
step (b) follows from the boundedness of
$w_{a}(\cdot)$ and that $\Delta - B\calE \ge 0$.
Taking $t = \frac{2}{\gamma \hat{\gamma}^2}(\Delta - B\calE)$, we have
\begin{align*}
\eqref{eq:eq23} =
\exp\Big(-\frac{1}{\gamma^2\hat{\gamma}^2}\cdot (\Delta - B\calE)^2|\calI_2| \Big) \le \delta,
\end{align*}
where the last inequality is due to the choice of $\Delta$.
As a result, we have with probability at least $1-\delta$
that $\hat{\alpha}(a(\alpha+\Delta)+\eps) > \alpha$, which implies 
that $a(\alpha+\Delta)+\eps > \hat{a}$. That is,
\$
\PP\big(a(\alpha+\Delta)+\eps > \hat a\big) \ge 1-\delta.
\$
Again, taking $\eps \rightarrow 0$ and using
the continuity of probability measures, we 
have w.p.~at least $1-\delta$ that
$a(\alpha+\Delta)\ge \hat{a}$, and thus complete the proof.
\end{proof}

\end{document}